\documentclass[twocolumn,superscriptaddress,nofootinbib
]{revtex4-1}

\renewcommand{\thetable}{\arabic{table}}

\makeatletter
\let\@fnsymbol\@arabic
\makeatother

\usepackage{graphicx}% Include figure files
\usepackage{amssymb}
\usepackage{dcolumn}% Align table columns on decimal point
\usepackage{bm}% bold math
\usepackage{siunitx}
\usepackage{mhchem}
\usepackage{longtable}
\usepackage{tabu}
\usepackage{nicefrac}

\usepackage{textpos}
\usepackage{hyperref}
\usepackage{amsmath}
\usepackage{breqn}
\usepackage{wasysym}
\usepackage{longtable}
\usepackage{supertabular}
\usepackage{natbib}
\setcitestyle{numbers,super}

\usepackage[symbol]{footmisc}

\renewcommand{\thefootnote}{\fnsymbol{footnote}}

\newcommand{\beginsupplement}{%
        \setcounter{table}{0}
        \renewcommand{\thetable}{S\arabic{table}}%
        \setcounter{figure}{0}
        \renewcommand{\thefigure}{S\arabic{figure}}%
        \setcounter{equation}{0}
        \renewcommand{\theequation}{S\arabic{equation}}%
     }

\makeatletter
\let\cat@comma@active\@empty
\makeatother

\begin{document}

\title{A Consistent Reduced Network for HCN Chemistry in Early Earth and Titan Atmospheres: Quantum Calculations of Reaction Rate Coefficients}

\author{Ben K. D. Pearce}
\email[Corresponding author:]{pearcbe@mcmaster.ca}
\affiliation{Origins Institute and Department of Physics and Astronomy, McMaster University, ABB 241, 1280 Main St, Hamilton, ON, L8S 4M1, Canada}
\author{Paul W. Ayers}
\affiliation{Origins Institute and Department of Chemistry and Chemical Biology, McMaster University, ABB 156, 1280 Main St, Hamilton, ON, L8S 4M1, Canada}
\author{Ralph E. Pudritz}
\affiliation{Origins Institute and Department of Physics and Astronomy, McMaster University, ABB 241, 1280 Main St, Hamilton, ON, L8S 4M1, Canada}

\begin{textblock*}{100mm}(-10cm,-1.1cm)
Submitted: 22 Nov 2018
\end{textblock*}

\begin{textblock*}{100mm}(-2.5cm,-1.1cm)
Accepted: 4 Feb 2019
\end{textblock*}

\begin{textblock*}{100mm}(4cm,-1.1cm)
doi: \href{http://dx.doi.org/10.1021/acs.jpca.8b11323}{10.1021/acs.jpca.8b11323}
\end{textblock*}

\begin{abstract}
{\bf Abstract:} HCN is a key ingredient for synthesizing biomolecules such as nucleobases and amino acids.
We calculate 42 reaction rate coefficients directly involved with or in competition with the production of HCN in the early Earth or Titan atmospheres. These reactions are driven by methane and nitrogen radicals produced via UV photodissociation or lightning. For every reaction in this network, we calculate rate coefficients at 298 K using canonical variational transition state theory (CVT) paired with computational quantum chemistry simulations at the BHandHLYP/aug-cc-pVDZ level of theory. We also calculate the temperature dependence of the rate coefficients for the reactions that have barriers from 50--400 K. We present 15 new reaction rate coefficients with no previous known value. 93$\%$ of our calculated coefficients are within an order of magnitude of the nearest experimental or recommended values. Above 320 K, the rate coefficient for the new reaction \ce{H2CN -> HCN + H} dominates. Contrary to experiments, we find the HCN reaction pathway, \ce{N + CH3 -> HCN + H2}, to be inefficient, and suggest the experimental rate coefficient actually corresponds to an indirect pathway, through the H$_2$CN intermediate. We present CVT using energies computed with density functional theory as a feasible and accurate method for calculating a large network of rate coefficients of small-molecule reactions.
\end{abstract}

\maketitle

\section*{Introduction}

HCN is a precursor to the building blocks of life. For example, HCN reacts to produce nucleobases, the building blocks of RNA/DNA, as well as amino acids, the building blocks of proteins, in aqueous environments \cite{Reference437,1961Natur.191.1193O,2008OLEB...38..383L,2014ApJ...783..140C,Reference46}. For adenine synthesis, HCN first condenses in water to form oligomers, which then forms adenine upon hydrolysis \cite{Reference12}. HCN may have formed in the atmosphere of the prebiotic Earth through the reaction of photochemically driven and/or lightning-induced methane and nitrogen radicals \cite{2007AsNow..22e..76R,Reference440}. HCN is similarly produced in Titan's present-day atmosphere \cite{2015Icar..247..218L}.

Given the significance HCN as a precursor to biomolecules, it is of interest to discern how much was produced in the early Earth atmosphere in order to understand whether it potentially played a role in the emergence of life in warm little ponds \cite{Reference87}. Titan provides a good test environment for atmospheric HCN production, given that one can compare abundances from chemical simulations to the measured HCN profile from the Cassini mission \cite{2011Icar..214..584A,2010Icar..205..559V}.

Chemical networks including a variety of species and reactions have been employed to simulate the atmospheric HCN composition of early Earth \cite{Reference591,2011EPSL.308..417T} and Titan \cite{2008PSS...56...27L,2008PSS...56...67L,2009Icar..201..226K,2012AA...541A..21H,2015Icar..247..218L}. The reaction rate coefficients in these networks are generally a combination of a) theoretical, b) experimental, and c) suggested values typically estimated using thermodynamics, similar reactions and/or experimental results at much higher temperatures. Each of these sources has errors associated with it, and there are often a range of experimental and theoretical values to choose from for a single reaction. As a result, atmospheric HCN compositions can vary by orders of magnitude from one simulation to the next. Therefore, it is perhaps unsurprising that, as of yet, no simulation has matched the HCN profile of Titan completely.

There are also several reactions without past experimental, theoretical, or suggested values that are missing in these networks that may play important roles in HCN formation (e.g. \ce{^1CH2 + ^2N -> H2CN} and \ce{H2CN -> HCN + H}).

The focus of this work is to create a theoretical reduced HCN chemical network, where all the rate coefficients are consistently calculated with the same theoretical and computational method. Using this strategy, all reactions can be theoretically validated before being employed in a chemical network, and key reaction pathways with previously unknown rate coefficients can be included. Furthermore, by constructing a model chemistry \cite{Reference592,Reference593} the errors for consistently calculated rate coefficients are expected to be similar, thus employing such a network has a chance to improve accuracy.

The limitation of calculating a consistent theoretical network is that one cannot feasibly include a large number of molecular species. For every additional species, there is a potential additional reaction with all the existing species in the network. Therefore in this work, we focus only on the small set of reactions involved in the production of HCN from methane and nitrogen dissociation radicals, as well as the direct competing reactions. This totals 42 reactions between 11 species. We are the first to calculate a completely consistent theoretical reaction network of this size for atmospheric chemistry simulations.

In the Background section of this paper, we motivate and describe the reactions in our chemical network. Then in the Methods section, we detail the theoretical and computational methods used to calculate the reaction rate coefficients in our network. In the Results section, we present the results of our calculations, including their conformance to experimental values, and the effects of spin configuration on these values. The reader who is just interested in the calculated rate coefficients can skip ahead to Tables~\ref{Table3} and \ref{TableTempDep}, where we present the calculated reaction rate coefficients at 298 K, and the Arrhenius coefficients for temperature dependences, respectively. Finally, in the Conclusions section we summarize the main results of the paper.

The supporting information (SI) contains a wealth of technical data and calculation details including: 1) a summary of the experimentally measured and previously theoretically calculated rate coefficients in this network, 2) an example rate coefficient calculation using the CVT method and a computational methods comparison, 3) a breakdown of the calculations of specific reactions, and 4) reaction path symmetry number calculations.

%For example, a simulation of cold dense clouds comparing the KIDA and UMIST astrochemical databases led to several-order-of-magnitude differences in equilibrium abundances for certain molecules \cite{2013ChRv..113.8710A}.

%We use canonical variational transition state theory (CVT) and computational quantum simulations to calculate a consistent reduced chemical network for producing HCN in an atmosphere.

\section*{Background}\label{background}

The abiotic production of biomolecules such as nucleobases and amino acids requires a reactive source of nitrogen, typically HCN or NH$_3$ \cite{Reference46,Reference438,Reference44,Reference440,2013NatGe...6.1045M}. HCN can be produced in early Earth and Titan atmospheres through reactions involving N$_2$ and CH$_4$ dissociation products. Such dissociation products are produced when N$_2$ and CH$_4$ interact with UV photons \cite{2007AsNow..22e..76R}, cosmic rays \cite{Reference596}, or lightning \cite{2014Icar..237..182C}.
N$_2$ and CH$_4$ photodissociation can be broken down into the following pathways

\begin{equation}\label{Eq1a}
\ce{N2 + $h \nu$ -> N2^+ + e^-}
\end{equation}
\begin{equation}\label{Eq1b}
\ce{N2^+ + e^- -> ^4N + ^2N}
\end{equation}
\begin{multline}\label{Eq2}
\ce{CH4 + $h \nu$ -> CH3 + H} \\ \Phi_{118.2} = 0.26, \Phi_{121.6} = 0.42,
\end{multline}
\begin{multline}\label{Eq3}
\ce{CH4 + $h \nu$ -> ^1CH2 + H2} \\ \Phi_{118.2} = 0.17, \Phi_{121.6} = 0.48,
\end{multline}
\begin{multline}\label{Eq4}
\ce{CH4 + $h \nu$ -> ^3CH2 + 2H} \\ \Phi_{118.2} = 0.48, \Phi_{121.6} = 0.03,
\end{multline}
\begin{multline}\label{Eq5}
\ce{CH4 + $h \nu$ -> CH + H2 + H} \\ \Phi_{118.2} = 0.09, \Phi_{121.6} = 0.07,
\end{multline}
where the leading superscripts signify the singlet, doublet, and quartet spin states, $h \nu$ signifies an ultraviolet photon and $\Phi_{118.2}$ and $\Phi_{118.2}$ signify the branching ratios measured from lab experiments at 118.2 and 121.6 nm, respectively \cite{2007AsNow..22e..76R,2017ApJ...850...48S,Reference586}.

Multiple possible pathways to produce HCN from the above radicals (at or near 298 K) have been reported from experiments or suggested in the literature. Note that molecular spin states are not included in this list and that each of these reactions represents 1--5 reaction spin configurations; each with a unique reaction rate coefficient.

\begin{equation}\label{Eq6}
^{\dagger} \ce{CH3 + N -> H2CN + H}
\end{equation}
\begin{equation}\label{Eq7}
^{\dagger} \ce{CH2 + N -> H2CN}
\end{equation}
\begin{equation}\label{Eq8}
^{\dagger} \ce{H2CN <-> HCN + H}
\end{equation}
\begin{equation}\label{Eq9}
\ce{H2CN + H -> HCN + H2} 
\end{equation}
\begin{equation}\label{Eq10}
\ce{H2CN + N -> HCN + NH}
\end{equation}
\begin{equation}\label{Eq11}
\ce{2H2CN -> HCN + H2CNH} 
\end{equation}

\renewcommand*{\thefootnote}{\fnsymbol{footnote}}

\footnotetext[2]{Reactions without experimental or suggested values for at least one spin configuration in this network.}

\renewcommand*{\thefootnote}{\arabic{footnote}}

\begin{table*}[t]
\centering
\caption{List of primary molecular species involved in this study and their spin states. \label{Table1}} 
\begin{tabular}{ccc}
\\
\multicolumn{1}{c}{Species} & 
\multicolumn{1}{c}{\hspace{0.3cm}Spin state} & 
\multicolumn{1}{c}{\hspace{0.3cm}Ground/Excited state}\\ \hline \\[-2mm]
HCN & singlet & ground \\
H$_2$CN & doublet & ground \\
N$_2$ & singlet & ground \\
$^2$N & doublet & excited \\
$^4$N & quartet & ground \\
CH$_4$ & singlet & ground \\
CH$_3$ & doublet & ground \\
$^1$CH$_2$ & singlet & excited \\
$^3$CH$_2$ & triplet & ground \\
CH & doublet & ground \\
H$_2$ & singlet & ground \\
H & doublet & ground \\
$^3$NH & triplet & ground \\
\hline
\end{tabular}
\end{table*}

Three experimentally reported or suggested reaction pathways have not been included in this list as our theoretical work shows they more likely proceed through two steps involving combinations of the above equations. These reactions are \ce{CH3 + N -> HCN + H2} \cite{Reference442}, \ce{CH3 + N -> HCN + 2H} \cite{Reference442}, and \ce{CH2 + N -> HCN + H} \cite{2007AsNow..22e..76R} (see theoretical case studies in SI for complete analysis).

There are also multiple competing reaction pathways to the above reactions at or near 298 K. In this network, we only include competing pathways involving the radicals produced from N$_2$ and CH$_4$ dissociation in the atmosphere. One exception is that we also include the reactions of $^3$NH with H and N as recombination pathways to H$_2$ and N$_2$. See Table~\ref{Table1} for list of primary molecular species.

\begin{equation}\label{Eq12}
\ce{CH4 + H <-> CH3 + H2}
\end{equation}
\begin{equation}\label{Eq13}
\ce{CH4 + N -> H2CNH + H}
\end{equation}
\begin{equation}\label{Eq14}
\ce{CH3 + H -> CH4}
\end{equation}
\begin{equation}\label{Eq15}
\ce{2CH3 -> C2H6}
\end{equation}
\begin{equation}\label{Eq16}
\ce{CH2 + H -> CH3}
\end{equation}
\begin{equation}\label{Eq17}
\ce{CH2 + H2 -> CH4}
\end{equation}
\begin{equation}\label{Eq18}
^{\dagger} \ce{CH2 + H2 <-> CH3 + H}
\end{equation}
\begin{equation}\label{Eq19}
\ce{2CH2 -> C2H4}
\end{equation}
\begin{equation}\label{Eq20}
\ce{CH2 + CH3 -> C2H4 + H}
\end{equation}
\begin{equation}\label{Eq21}
\ce{CH2 + CH4 -> C2H6}
\end{equation}
\begin{equation}\label{Eq22}
^{\dagger} \ce{CH2 + CH4 <-> 2CH3}
\end{equation}
\begin{equation}\label{Eq23}
^{\dagger} \ce{CH + H -> CH2}
\end{equation}
\begin{equation}\label{Eq24}
\ce{CH + H2 -> CH3}
\end{equation}
\begin{equation}\label{Eq25}
^{\dagger} \ce{CH + N -> CN + H}
\end{equation}
\begin{equation}\label{Eq26}
\ce{2CH -> C2H2}
\end{equation}
\begin{equation}\label{Eq27}
\ce{CH + CH4 -> C2H4 + H}
\end{equation}
\begin{equation}\label{Eq28}
\ce{NH + H <-> H2 + N}
\end{equation}
\begin{equation}\label{Eq29}
^{\dagger} \ce{NH + N -> N2 + H}
\end{equation}

Four experimentally reported \cite{Reference503,Reference504,Reference505,Reference506,Reference507,Reference508,Reference517,Reference510,Reference514,Reference515} 
 two-step reaction pathways have been reduced to their first steps in this list. These reactions are 

\begin{equation*}
\ce{CH + H2 <-> CH3* <-> ^{3,1}CH2 + H},
\end{equation*}
\begin{equation*}
\ce{^1CH2 + H2 -> CH4* -> CH3 + H},
\end{equation*}
\begin{equation*}
\ce{^1CH2 + CH4 -> C2H6* -> 2CH3}.
\end{equation*}
 
Our theoretical work shows the first steps are the rate-limiting steps, and the intermediates are reactants with other available reaction pathways in our chemical network (see theoretical case studies in SI for complete details).

One other experimentally reported \cite{Reference580} reaction has not been included in this list. This reaction is

\begin{equation*}
\ce{CH4 + ^2N -> ^1H3CNH* -> CH3 + ^3NH}.
\end{equation*}

Experiments suggest that $^1$H$_3$CNH decays into CH$_3$ + $^3$NH with a branching ratio of 0.3 $\pm$ 0.1, and that the majority of $^1$H$_3$CNH decays into $^1$H$_2$CNH + H ($\Phi$ = 0.8 $\pm$ 0.2). Our theoretical work also suggests $^1$H$_3$CNH preferentially decays into $^1$H$_2$CNH + H, however we alternatively find the decay into CH$_3$ + $^3$NH to be very inefficient (k $\sim$ 10$^{-29}$ cm$^3$ s$^{-1}$); therefore we do not consider this decay pathway in this network.

The focus of this work is to calculate the rate coefficients for an atmospheric HCN reaction network which can be applied to both Titan and early Earth atmospheres. For each reaction, a detailed analysis of spin state configurations and an series of computational quantum chemistry simulations are performed at temperatures between 50--400 K.

In Table~\ref{Table2} we summarize the molecules and spin states involved in this reaction network. We define reactions with rate coefficients greater than 10$^{-21}$ s$^{-1}$ for unimolecular reactions or greater than 10$^{-21}$ cm$^{3}$s$^{-1}$ for bimolecular reactions as ``fast,'' and exclude the ``slow'' reactions with smaller rate coefficients from this network.

\begin{table*}[!ht]
\centering
\caption{Detailed list of reactions considered in this study, including the accessible potential energy surfaces, and spin-state configurations. The focus of this network is reactions involved in the production of HCN from nitrogen and methane dissociation radicals. Direct competing reactions are also included. We define a fast reaction rate coefficient to be $>$10$^{-21}$ s$^{-1}$ for unimolecular reactions and $>$10$^{-21}$ cm$^{3}$s$^{-1}$ for bimolecular reactions. \label{Table2}}
\resizebox{\textwidth}{!}{% 
\begin{tabular}{ccccc}
\\
\multicolumn{1}{c}{Reaction equation} & 
\multicolumn{1}{c}{PES} & 
\multicolumn{1}{c}{Spin Configuration} &
\multicolumn{1}{c}{Fast k$_f$(298)?} &
\multicolumn{1}{c}{Fast k$_r$(298)?}\\ \hline \\[-2mm]
\ce{H2CN <-> HCN + H} &	doublet & \ce{H2CN <-> HCN + H} & Y & Y \\
\ce{H2CN + H <-> HCN + H2} &	singlet & \ce{H2CN + H <-> HCN + H2} & Y & N \\
\ce{H2CN + N <-> HCN + NH} & singlet & \ce{H2CN + ^2N <-> HCN + ^1NH} & N & N \\
 & triplet & \ce{H2CN + ^4N <-> HCN + ^3NH} & Y & N \\
  &  & \ce{H2CN + ^2N <-> HCN + ^3NH} & N & N \\
\ce{2H2CN <-> HCN + H2CNH} & singlet & \ce{2H2CN <-> HCN + H2CNH} & Y & N \\
\ce{CH4 + H <-> CH3 + H2} & doublet & \ce{CH4 + H <-> CH3 + H2} & Y & Y \\
\ce{CH4 + N <-> H3CNH* <-> H2CNH + H} & doublet & \ce{CH4 + ^2N <-> H3CNH* <-> H2CNH + H} & Y & N \\
\ce{CH4 + N <-> H3CNH* <-> CH3 + NH} & doublet & \ce{CH4 + ^2N <-> H3CNH* <-> CH3 + ^3NH} & N & N \\
\ce{CH3 + H <-> CH4} & singlet & \ce{CH3 + H <-> CH4} & Y & N \\
\ce{CH3 + N <-> H3CN* <-> HCN + H2} &	singlet & \ce{CH3 + ^2N <-> ^1H3CN* <-> HCN + H2} & N & N \\
\ce{CH3 + N <-> H3CN* <-> H2CN + H} &	singlet & \ce{CH3 + ^2N <-> ^1H3CN* <-> H2CN + H} & Y & N \\
 &	triplet & \ce{CH3 + ^4N <-> ^3H3CN* <-> H2CN + H} & Y & N \\
 &	 & \ce{CH3 + ^2N <-> ^3H3CN* <-> H2CN + H} & Y & N\\
\ce{2CH3 <-> C2H6} & singlet & \ce{2CH3 <-> C2H6} & Y & N  \\
\ce{CH2 + H <-> CH3} & doublet & \ce{^3CH2 + H <-> CH3} & Y & N \\
 &  & \ce{^1CH2 + H <-> CH3} & Y & N  \\
\ce{CH2 + H2 <-> CH4} & singlet & \ce{^1CH2 + H2 <-> CH4} & Y & N \\
\ce{CH2 + H2 <-> CH3 + H} & triplet & \ce{^3CH2 + H2 <-> CH3 + H} & Y & Y \\
\ce{CH2 + N <-> H2CN} &	doublet & \ce{^3CH2 + ^4N <-> ^2H2CN} & Y & N \\
 &	 & \ce{^3CH2 + ^2N <-> ^2H2CN} & Y & N \\
 &	 & \ce{^1CH2 + ^2N <-> ^2H2CN} & Y & N \\
&	quartet & \ce{^1CH2 + ^4N <-> ^4H2CN} & Y & N \\
&	 & \ce{^3CH2 + ^2N <-> ^4H2CN} & Y & N \\
\ce{2CH2 <-> C2H4} & singlet & \ce{^3CH2 + ^3CH2 <-> ^1C2H4} & Y & N  \\
 &  & \ce{^1CH2 + ^1CH2 <-> ^1C2H4} & Y & N \\
 & triplet & \ce{^3CH2 + ^1CH2 <-> ^3C2H4} & Y & N  \\
 \ce{CH2 + CH3 <-> C2H5* <-> C2H4 + H} & doublet & \ce{^3CH2 + CH3 <-> C2H5* <-> ^1C2H4 + H} & Y & N \\
 &  & \ce{^1CH2 + CH3 <-> C2H5* <-> ^1C2H4 + H} & Y & N  \\
  & quartet & \ce{^3CH2 + CH3 <-> ^4C2H5* <-> ^3C2H4 + H} & N & N/A \\
\ce{CH2 + CH4 <-> C2H6} & singlet & \ce{^1CH2 + CH4 <-> C2H6} & Y & N  \\
\ce{CH2 + CH4 <-> 2CH3} & triplet & \ce{^3CH2 + CH4 <-> 2CH3} & Y & Y  \\
\ce{CH + H <-> CH2} & singlet & \ce{CH + H <-> ^1CH2} & Y & N  \\
 & triplet & \ce{CH + H <-> ^3CH2} & Y & N  \\
\ce{CH + H2 <-> CH3} & doublet & \ce{CH + H2 <-> CH3} & Y & N  \\
\ce{CH + N <-> HCN <-> CN + H} & triplet & \ce{CH + ^4N <-> ^3HCN <-> CN + H} & Y & N  \\
& & \ce{CH + ^2N <-> ^3HCN <-> CN + H} & Y & N  \\
\ce{2CH <-> C2H2} & singlet & \ce{CH + CH <-> C2H2} & Y & N  \\
\ce{CH + CH4 <-> C2H5* <-> C2H4 + H} & doublet & \ce{CH + CH4 <-> C2H5* <-> C2H4 + H} & Y & N  \\	
\ce{NH + H <-> H2 + N} & doublet & \ce{^1NH + H <-> H2 + ^2N} & N/A & N \\
 &  & \ce{^3NH + H <-> NH2* <-> H2 + ^2N} & N & Y \\
 & quartet & \ce{^3NH + H <-> H2 + ^4N} & Y & N \\
\ce{NH + N <-> N2H* <-> N2 + H} & doublet & \ce{^3NH + ^4N <-> N2H* <-> N2 + H} & Y & N  \\
 &  & \ce{^1NH + ^2N <-> N2H* <-> N2 + H} & N/A & N \\
 &  & \ce{^3NH + ^2N <-> N2H* <-> N2 + H} & Y & N \\
\hline
\multicolumn{5}{l}{\footnotesize Reactions are N/A if they require species that are not efficiently produced in this network.} \\
\end{tabular}}
\end{table*}

\section*{Methods}\label{methods}

\subsection*{Variational Transition State Theory}

\begin{figure}[hbtp]
\centering
\includegraphics[width=\linewidth]{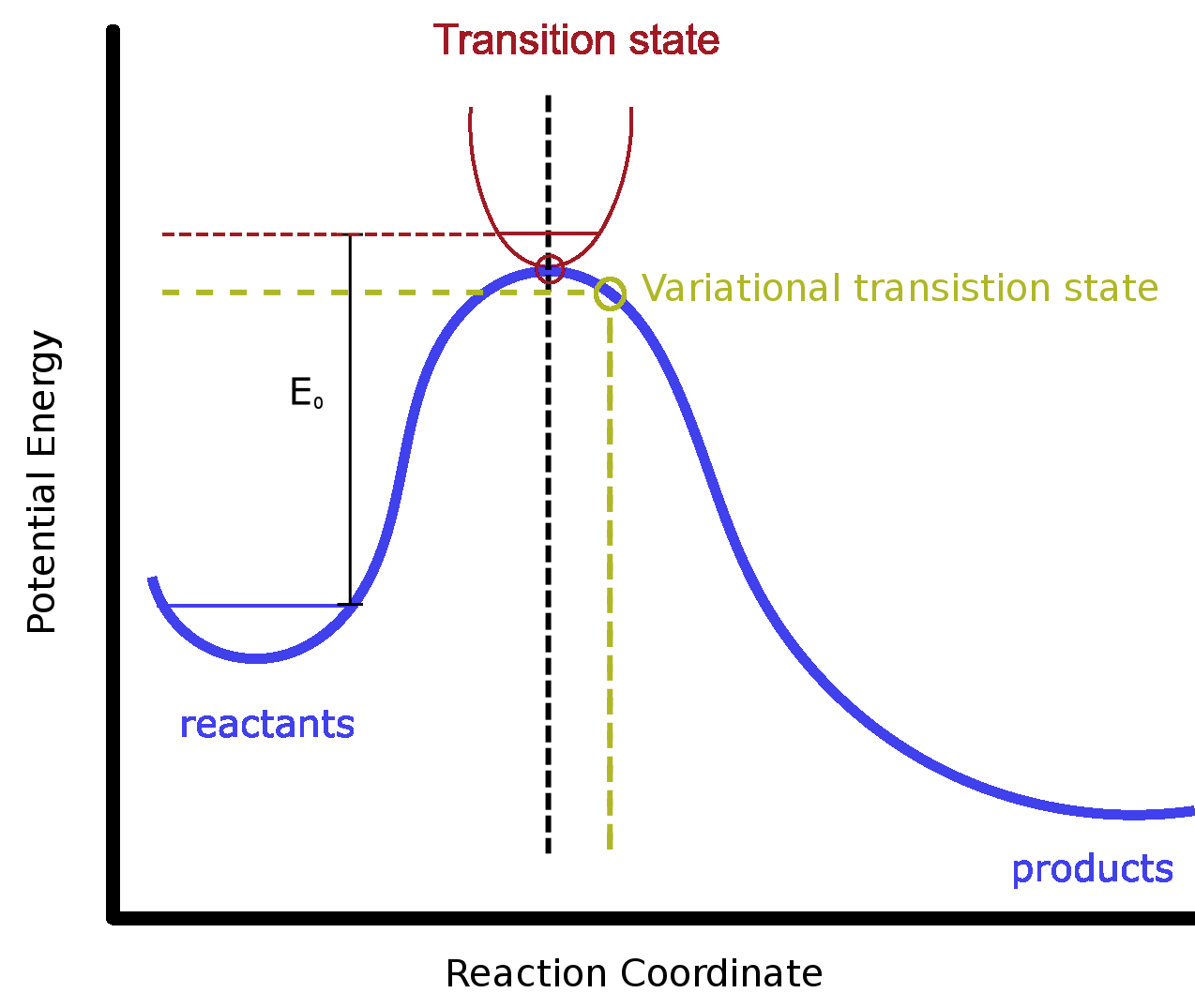}
\caption{A schematic representation of a reaction: proceeding from the reactants, over the potential energy barrier, E$_0$, through the transition state (red circle), and onto the products. The variational transition state (gold circle) is a location beyond the conventional transition state, where reactants that recross the barrier tend not to reach. The variational transition state is located where the reaction rate coefficient is at a minimum, thus providing the best dynamical bottleneck.}
\label{Figure1}
\end{figure}

Reactions can be visualized in one dimension using potential energy diagrams (see Figure~\ref{Figure1}). A reaction proceeds along a coordinate (e.g. the distance between two atoms), from the reactant geometry, to the product geometry. In some cases, the minimum energy path (MEP) from reactants to products requires proceeding through a geometry of higher potential energy than the reactant and product geometries. This increase in potential energy along a reaction coordinate is known as the energy barrier. The peak of the energy barrier describes the conventional transition state.

In reality, reactions have more than one dimension (e.g. bond distances, angles between bonds, dihedral angles), thus the energy barrier is more appropriately described as a saddle point, and the MEP is the path of steepest descent from saddle point to the reactant and product minima. The rate of a reaction can be described as how frequently molecules travel the entire MEP, and is quantified by the reaction rate coefficient, $k$.

We calculate gas phase chemical reaction rate coefficients using canonical variational transition state theory (CVT). The basis for this method is to vary the reaction coordinate (e.g. the carbon-hydrogen bond distance) along the MEP in order to find the minimum rate constant. Unlike conventional transition state theory, CVT allows us to calculate reaction rate coefficients for both barrierless and non-barrierless reactions, while minimizing the error due to trajectories that recross the transition state rather than descend into products \cite{Reference534}. This can be visualized as finding a location past the saddle point of the MEP, that recrossing reactants tend not to reach (see Figure~\ref{Figure1}). This location is determined as the location where the generalized transition state (GT) rate coefficient is at its smallest value, therefore providing best dynamical bottleneck \cite{Reference534}.

%The reactions we are interested are bimolecular (e.g. recombination, displacement, synthesis) with units cm$^3$s$^{-1}$, and unimolecular (decomposition) with units s$^{-1}$.

The CVT reaction rate coefficient is expressed as \cite{Reference535,Reference456,Reference495}

\begin{equation}\label{CVT}
k_{CVT}(T,s) = \min_s \left\lbrace k_{GT}(T,s) \right\rbrace.
\end{equation}

Neglecting the tunneling effect, the generalized transition state theory (GT) reaction rate coefficient can be approximated via the Eyring Equation \cite{Reference530,Reference535,Reference532,Reference531,Reference529,Reference456,Reference536,Reference495,Reference537}. The Eyring equation uses a statistical mechanics approach to calculate the rate coefficient by dividing the density of forward-crossing states per unit time by the density of reactant states.

\begin{equation}\label{Eyring}
k_{GT}(T,s) = \sigma \frac{k_B T}{h} \frac{Q^{\ddagger}(T,s)}{\prod_{i=1}^{N} Q_i^{n_i}(T)} e^{-E_0(s)/RT}
\end{equation}
where $\sigma$ is the reaction path symmetry number or reaction path multiplicity (i.e. the number of equivalent reaction paths from reactants to products), {\it k$_B$} is the Boltzmann constant (1.38$\times$10$^{-23}$ J K$^{-1}$), {\it T} is temperature (K), {\it h} is the Planck constant (6.63$\times$10$^{-34}$ J$\cdot$s), {\it Q$^{\ddagger}$} is the partition function of the transition state per unit volume (cm$^{-3}$), with its zero of energy at the saddle point, {\it Q$_i$} is the partition function of species $i$ per unit volume, with its zero of energy at the equilibrium position of species $i$ (i.e. as if it is infinitely separated from any other reactant), $n_i$ is the stoichiometric coefficient of species $i$, $N$ is the number of reactant species, {\it E$_0$} is the energy barrier (the difference in zero-point energies between the generalized transition state and the reactants) (kJ mol$^{-1}$), and {\it R} is the gas constant (8.314$\times$10$^{-3}$ kJ K$^{-1}$ mol$^{-1}$).

Because classical partition functions involve integrating over the Boltzmann factor ($e^{-E/RT}$), an additional exponential factor appears naturally in the Eyring equation due to the difference in zeros of energy between the transition state and reactant states.

To find the location along the MEP where the GT rate coefficient is at its smallest value, we use the maximum Gibbs free energy criterion, which gives a compromise of energetic and entropic effects \cite{Reference535,Reference538}. To obtain a similar accuracy for all calculations, we use a reaction coordinate precision of 0.01 $\AA$. Looking at the quasithermodynamic representation of transition-state theory, we see that the maximum value for $\Delta G_{GT}(T,s)$ corresponds to a minimum value for $k_{GT}(T,s)$

\begin{equation}\label{GibbsCrit}
k_{GT}(T,s) = \frac{k_B T}{h} K^0 e^{-\Delta G_{GT}(T,s)/RT},
\end{equation}
where $K^0$ is the reaction quotient under standard state conditions (i.e. 1 for unimolecular reactions, and 1 cm$^3$ for bimolecular reactions), and $\Delta G_{GT}(T,s)$ is the difference in the Gibbs free energy between transition state and reactants (kJ mol$^{-1}$).

The conventional transition state, energy barrier, and variational transition state are illustrated with a potential energy diagram in Figure~\ref{Figure1}. 

The zero-point energies and partition functions for the reactants and transition states are calculated using the Gaussian 09 software package \cite{g09}. A brief summary of the theory behind these calculations is detailed below. We refer the reader to Ochterski\cite{Reference533} for further details.

The partition functions per unit volume are expanded into their 4 components

\begin{equation}
Q = \frac{q_t}{V} q_e q_v q_r.
\end{equation}
where {\it q$_t$} is the translational component, V is the volume (cm$^{-3}$), {\it q$_e$} is the electronic component, {\it q$_v$} is the vibrational component, {\it q$_r$} is the rotational component not including the rotational symmetry number (this is included in the reaction path multiplicity).

From classical statistical mechanics, the translational partition function per unit volume is \cite{Reference533}

\begin{equation}
\frac{q_t}{V} = \left( \frac{2 \pi m k_B T}{h^2} \right)^{3/2},
\end{equation}
where {\it m} is the mass of the species (kg).

The electronic partition function is estimated as the degeneracy of the first energy level, {\it i.e.} the spin multiplicity \cite{Reference533}

\begin{equation}
q_e = 2S + 1,
\end{equation}

where S is the total spin due to unpaired electrons. For example, a hydrogen atom has 1 unpaired electron of spin $\nicefrac{1}{2}$, and thus its {\it q$_e$} = 2($\nicefrac{1}{2}$) + 1 = 2.

Gaussian calculates the vibrational partition function as a quantum harmonic oscillator. We note that for the zero-point energies of molecules, Gaussian places the zero of energy at the bottom of the internuclear potential. Thus, with this same location for the zero of energy, the vibrational partition function equates to

\begin{equation}
q_v = \prod_{n=1}^{N} \frac{e^{-\Theta_n/{2T}}}{1 - e^{-\Theta_n/T}},
\end{equation}
where $N$ is the number of vibrational modes, $\Theta_n$ is the vibrational temperature of the n$^{th}$ mode ($\Theta_n$ = $\frac{\hslash \omega_n}{k_B}$), and T is temperature. 

By default, Gaussian calculates the rotational partition function as a rigid rotor. For linear molecules excluding rotational symmetry,

\begin{equation}
q_r = \left( \frac{T}{\Theta_r} \right),
\end{equation}

and for polyatomic molecules excluding rotational symmetry,

\begin{equation}
q_r = \pi^{1/2}\left( \frac{T^{3/2}}{(\Theta_{r,x} \Theta_{r,y} \Theta_{r,z})^{3/2}} \right),
\end{equation}
where $\Theta_r$ is the rotational temperature ($\Theta_r = \frac{h^2}{8 \pi^2 I k_B}$, and $I$ is the moment of inertia (in the case of a polyatomic molecule, $I_x$, $I_y$, and $I_z$ are the principal moments of inertia).

Gaussian displays an output for the rotational symmetry number ($\sigma_r$) of each molecule, however for all the reactants, transition states and products in our study, Gaussian displayed $\sigma_r$ = 1. For this reason we calculate the rotational symmetry in Equation~\ref{Eyring} manually \cite{Reference536} (the calculated symmetry numbers are listed in Table S10 in SI).

\subsection*{Quantum Computational Simulations}

We perform quantum computational simulations with the Gaussian software package \cite{g09} using the Becke-Half-and-Half-Lee-Yang-Parr (BHandHLYP) density functional \cite{Reference594,Reference595}. We chose BHandHLYP for two reasons. Firstly, it is a relatively inexpensive method that can be used for an extended transition state study such as this. Secondly, in a computational methods comparison of the well-studied reaction \ce{CH4 + H -> CH3 + H2}, BHandHLYP provided the most accurate rate coefficient compared to calculations using HF, CCSD, B3LYP, and M06-2x (see computational methods comparison in SI for more details). CAM-B3LYP also provided an accurate rate coefficient for this reaction, however the value from BHandHLYP offers a better compromise between experimental and suggested values.

Hartree-Fock (HF) methods tend to overestimate the energy barrier, whereas Density Functional Theory (DFT) methods (e.g. B3LYP) tend to underestimate the energy barrier. BHandHLYP is a hybrid functional that improves performance by using 50$\%$ HF and 50$\%$ DFT for the exchange energy calculation. All simulations are performed with the augmented correlation consistent polarized valence double zeta (aug-cc-pVDZ) basis set in order to achieve reasonable computation times.

Typically, when there is only one reaction spin configuration for a given PES, we do not specify the local spins in Gaussian when calculating the MEP. However in some cases not specifying the local spin, regardless of the number of possible spin configurations, leads to convergence issues. In these cases we specify the local spins to allow the calculation to converge. When there is more than one reaction spin configuration for a given PES, e.g., \ce{CH3 + ^4N -> ^3H3CN* -> H2CN + H} and \ce{CH3 + ^2N -> ^3H3CN* -> H2CN + H} on the triplet surface, we specify the local spins of the reactants in Gaussian to find the MEP's for each individual spin configuration.

\subsection*{Temperature Dependence of Rate Coefficients}

Temperatures in the early Earth and Titan atmospheres fit comfortably within the range of 50--400 K \cite{2011EPSL.308..417T,2017JGRE..122..432H,Reference80}. The CVT rate coefficient equation for reactions with barriers includes a temperature-dependent exponential term (see Equation~\ref{Eyring}). This exponential temperature dependence typically leads to reaction rate coefficients which vary by multiple orders of magnitude over 50--400 K. The exponential term is omitted for barrierless reactions, and thus the temperature dependence for barrierless reaction rate coefficients is much smaller. Typically rate coefficients for barrierless reactions have either no temperature dependence, or a weak temperature dependence, varying by less than a factor of a two or three from 50--400 K \cite{Reference587,Reference495,Reference481,Reference577,Reference572}.

Temperature dependence for rate coefficients can be expressed using the Arrhenius equation \cite{2010AA...522A..42S},

\begin{equation}
k(T) = \alpha \left(\frac{T}{300}\right)^{\beta} e^{-\gamma/T},
\end{equation}
where $\alpha$, $\beta$, and $\gamma$ are fitting parameters, which we will refer to as the Arrhenius coefficients. Units for $k(T)$ are s$^{-1}$ for unimolecular reactions and cm$^{3}$s$^{-1}$ for bimolecular reactions.

We calculate the rate coefficients for the reactions with barriers at 50, 100, 200, 298, and 400 K and fit the results to the expression above to obtain the Arrhenius coefficients. For the sake of feasibility, we assume the rate coefficient for barrierless reactions is constant within this temperature range, as is typical \cite{Reference451}.

\section*{Results}\label{results}

For detailed results, see theoretical case studies for 35 of the reactions in SI.

%In Table~\ref{Table2} we display the spin configurations for each reaction in this study and state which forward and reverse spin configurations have rate coefficients greater than 10$^{-21}$ s$^{-1}$ or 10$^{-21}$ cm$^3$ s$^{-1}$.

%\ce{CH2 + N2 -> HCN + NH} & singlet & \ce{^1CH2 + N2 -> HCN + ^1NH} & No &  & \citet{Reference449}\\
% & triplet & \ce{^3CH2 + N2 <-> HCN + ^3NH} & No &  & \citet{Reference448}\\

In Table~\ref{Table3}, we display the reaction rate coefficients calculated using the CVT method described above at 298 K, and the comparative ranges of experimental values. 

\subsection*{Conformance to Experiments}

Of the 42 total reactions in this network, $\sim$54$\%$ have been studied experimentally at or near 298 K (see the ``k(298) experimental'' column in Table~\ref{Table3} for experimental values). Another $\sim$10$\%$ have been estimated based on the rate coefficients of similar bond additions and decompositions, and/or thermodynamics. 36$\%$ of the reactions have no experimental rate coefficients (those with no ``k(298) experimental'' value in Table~\ref{Table3}), and in most cases, we are the first to calculate them theoretically.

\setlength\LTcapwidth{\textwidth}
\begin{longtable*}{cccccc}
\caption{Reaction rate coefficients for the atmospheric reaction network calculated in this study. All reactions are involved in HCN production in the early Earth atmosphere or are key competing reactions. Calculations are performed at the BHandHLYP/aug-cc-pVDZ level of theory. Slow reactions (k $<$ 10$^{-21}$), either forward or reverse, are not included in this network. In the column labeled ``barrier?'' we specify whether the rate-limiting step (or the only step) of the reaction has an energy barrier. The error factor is the multiplicative or divisional factor from the nearest experimental or suggested value; the error factor is 1 if the calculated value is within the range of experimental or suggested values. 36$\%$ of these reactions have no experimental or suggested rate coefficients. First-order rate coefficients have units s$^{-1}$. Second-order rate coefficients have units cm$^{3}$s$^{-1}$. \label{Table3}} \\
Reaction equation & Forward or Reverse? & Barrier? & k(298) calculated & k(298) experimental & Error factor\\ \hline \\[-2mm]
\ce{H2CN <-> HCN + H} & F & Y & 1.6$\times$10$^{-11}$ & &\\
& R & Y & 2.7$\times$10$^{-14}$ & &\\
\ce{H2CN + H -> HCN + H2} & F & N & 1.8$\times$10$^{-11}$ & 8.3$\times$10$^{-11}$ & 5\\
\ce{H2CN + ^4N <-> HCN + ^3NH} & F & Y & 9.4$\times$10$^{-13}$ & 4.4$\times$10$^{-11}$ & 47 \\
\ce{2H2CN <-> HCN + H2CNH} & F & N & $^a$3.7$\times$10$^{-14}$ & 3.3--8.3$\times$10$^{-12}$ & 89 \\
\ce{CH4 + H <-> CH3 + H2} & F & Y & 8.1$\times$10$^{-18}$ & 8.2$\times$10$^{-19}$--3.5$\times$10$^{-17}$ & 1\\
 & R & Y & 3.2$\times$10$^{-21}$ & 9.6$\times$10$^{-21}$--1.3$\times$10$^{-20}$ & 3\\
\ce{CH4 + ^2N <-> H3CNH <-> ^1H2CNH + H} & F & Y & $^b$4.7$\times$10$^{-11}$ & 2.4--4.5$\times$10$^{-12}$ & 10\\
\ce{CH3 + H <-> CH4} & F & N & 7.9$\times$10$^{-11}$ & 1.5--4.7$\times$10$^{-10}$ & 2 \\
\ce{CH3 + ^4N <-> ^3H3CN* <-> H2CN + H} & F & N & 3.3$\times$10$^{-11}$ & 5.0--7.7$\times$10$^{-11}$ & 1.5 \\
\ce{CH3 + ^2N <-> ^3H3CN* <-> H2CN + H} & F & N & 1.0$\times$10$^{-10}$ & & \\
\ce{CH3 + ^2N <-> ^1H3CN* <-> H2CN + H} & F & N & 3.1$\times$10$^{-11}$ & & \\
\ce{2CH3 <-> C2H6} & F & N & 7.3$\times$10$^{-13}$ & 3.5--6.5$\times$10$^{-11}$ & 48\\
\ce{^1CH2 + H <-> CH3} & F & N & 8.4$\times$10$^{-11}$ & 5.0$\times$10$^{-11}$ & 2 \\
\ce{^1CH2 + H2 <-> CH4} & F & N & 1.0$\times$10$^{-11}$ & $^c$7.0$\times$10$^{-12}$--1.3$\times$10$^{-10}$ & 1 \\
\ce{^1CH2 + ^4N <-> ^4H2CN} & F & N & 1.1$\times$10$^{-10}$ & & \\
\ce{^1CH2 + ^2N <-> ^2H2CN} & F & N & 1.5$\times$10$^{-10}$ & & \\
\ce{^1CH2 + ^1CH2 <-> C2H4} & F & N & 9.9$\times$10$^{-12}$ & 5.0$\times$10$^{-11}$ & 5 \\
\ce{^1CH2 + ^3CH2 <-> C2H4} & F & N & $^d$3.5$\times$10$^{-11}$ & 3.0$\times$10$^{-11}$ & 1 \\
\ce{^1CH2 + CH3 <-> C2H5* <-> C2H4 + H} & F & N & 2.3$\times$10$^{-11}$ & 3.0$\times$10$^{-11}$ & 1 \\
\ce{^1CH2 + CH4 <-> C2H6} & F & N & 6.1$\times$10$^{-13}$ & $^e$1.9$\times$10$^{-12}$--7.3$\times$10$^{-11}$  & 3 \\
\ce{^3CH2 + H <-> CH3} & F & N & 5.6$\times$10$^{-10}$ & $^f$8.3$\times$10$^{-11}$--2.7$\times$10$^{-10}$ & 2 \\
\ce{^3CH2 + H2 <-> CH3 + H} & F & Y & 2.5$\times$10$^{-16}$ & $<$5.0$\times$10$^{-14}$--5.0$\times$10$^{-15}$ & $^g$ \\
 & R & Y & 1.4$\times$10$^{-20}$ & &  \\
\ce{^3CH2 + ^4N <-> ^2H2CN} & F & N & 1.3$\times$10$^{-10}$ & & \\
\ce{^3CH2 + ^2N <-> ^2H2CN} & F & N & 2.7$\times$10$^{-10}$ & & \\
\ce{^3CH2 + ^2N <-> ^4H2CN} & F & N & 4.3$\times$10$^{-10}$ & & \\
\ce{^3CH2 + ^3CH2 <-> C2H4} & F & N & 4.2$\times$10$^{-11}$ & 5.3$\times$10$^{-11}$ & 1 \\
\ce{^3CH2 + CH3 <-> C2H5* <-> C2H4 + H} & F & N & 8.8$\times$10$^{-12}$ & 5.0$\times$10$^{-11}$--2.1$\times$10$^{-10}$ & 6 \\
\ce{^3CH2 + CH4 <-> 2CH3} & F & Y & 1.4$\times$10$^{-16}$ & $<$5.0$\times$10$^{-14}$--3.0$\times$10$^{-19}$ & $^g$ \\
 & R & N & 5.5$\times$10$^{-11}$ &  & \\
\ce{CH + H <-> ^1CH2} & F & N & 1.5$\times$10$^{-10}$ &  &  \\
\ce{CH + H <-> ^3CH2} & F & N & 5.3$\times$10$^{-10}$ &  &  \\
\ce{CH + H2 <-> CH3} & F & N & 7.9$\times$10$^{-11}$ & 1.0$\times$10$^{-12}$--1.6$\times$10$^{-10}$ & 1 \\
\ce{CH + ^4N <-> ^3HCN <-> CN + H} & F & N & 1.1$\times$10$^{-10}$ & 2.1$\times$10$^{-11}$--1.6$\times$10$^{-10}$ & 1\\
\ce{CH + ^2N <-> ^3HCN <-> CN + H} & F & N & 2.7$\times$10$^{-10}$ & & \\
\ce{2CH <-> C2H2} & F & N & 1.3$\times$10$^{-11}$ & 1.7--2.0$\times$10$^{-10}$ & 13\\
\ce{CH + CH4 <-> C2H5* <-> C2H4 + H} & F & N & 3.8$\times$10$^{-13}$ & 2.0$\times$10$^{-12}$--3.0$\times$10$^{-10}$ & 5 \\
\ce{^3NH + H <-> H2 + ^4N} & F & Y & 1.4$\times$10$^{-11}$ & 3.2$\times$10$^{-12}$ & 4\\
\ce{^3NH + H <-> NH2* <-> H2 + ^2N} & R & Y & 5.1$\times$10$^{-11}$ & 1.7--5.0$\times$10$^{-12}$ & 10 \\
\ce{^3NH + ^4N <-> N2H* <-> N2 + H} & F & N & 4.0$\times$10$^{-11}$ & 2.5--2.6$\times$10$^{-11}$ & 1.5 \\
\ce{^3NH + ^2N <-> N2H* <-> N2 + H} & F & N & 5.5$\times$10$^{-11}$ & & \\
\hline
\multicolumn{6}{l}{\footnotesize $^a$ Simulations did not converge beyond a H-N bond distance of 1.95$\AA$. The calculated rate coefficient is an lower bound.} \\
\multicolumn{6}{l}{\footnotesize $^b$ Simulations did not converge beyond a H-N bond distance of 2.82$\AA$. The calculated rate coefficient is a lower bound.} \\
\multicolumn{6}{l}{\footnotesize $^c$ Experimental values are from the two-step reaction \ce{^1CH2 + H2 -> CH4* -> CH3 + H}. Our theoretical work suggests the} \\
\multicolumn{6}{l}{\footnotesize first step is the rate-limiting step, thus these values can be attributed to \ce{^1CH2 + H2 -> CH4}.} \\
\multicolumn{6}{l}{\footnotesize $^d$ Simulations did not converge beyond a C-C bond distance of 3.52$\AA$. The calculated rate coefficient is a lower bound.} \\
\multicolumn{6}{l}{\footnotesize $^e$ Experimental values are from the two-step reaction \ce{^1CH2 + CH4 -> C2H6* -> 2CH3}. Our theoretical work suggests the} \\
\multicolumn{6}{l}{\footnotesize first step is the rate-limiting step, thus these values can be attributed to \ce{^1CH2 + CH4 -> C2H6}.} \\
\multicolumn{6}{l}{\footnotesize $^f$ Experimental values are from the two-step reaction \ce{^3CH2 + H -> CH3* -> CH + H2}. Our theoretical work suggests the} \\
\multicolumn{6}{l}{\footnotesize first step is the rate-limiting step, thus these values can be attributed to \ce{^3CH2 + H -> CH3}.} \\

\multicolumn{6}{l}{\footnotesize $^g$ The theoretical value agrees with the experimental upper bounds.} \\
\end{longtable*}

It is often assumed that experiments provide the closest values to the true reaction rate coefficients. However for a single reaction, separate experiments can measure coefficients that differ by over 2 orders of magnitude (e.g. for \ce{CH + CH4 -> C2H4 + H}, $k$ = 2.0$\times$10$^{-12}$ to 3.0$\times$10$^{-10}$ cm$^{3}$s$^{-1}$). This variation can be due to differing experimental methods, instrumentation, and analytical techniques. Furthermore, the reactions reported in experiments may not correspond to direct pathways. Instead there may be intermediates embedded in multiple reaction steps that correspond to the overall measured reaction rate coefficient. Theoretical analysis and mechanistic modeling can be used to sort out the most likely steps in a multiple-step reaction in order to avoid the inclusion of redundant reaction pathways in chemical networks.

In this work, we calculate the reaction rate coefficients for the reactions involved in HCN production from atmospheric nitrogen and methane radicals, as well as the most direct competing reactions. This network includes 15 reactions that have no experimental or suggested value in the literature, and six of these are directly involved in atmospheric HCN synthesis. All our calculations are performed at the same level of theory, i.e. BHandHLYP/aug-cc-pVDZ, therefore we expect the error in the rate coefficients to be similar for all reactions.

The largest discrepancy between experiments and theory is for the reaction of \ce{CH3 + N -> products}. Stief et al.\cite{Reference444} measured the rate coefficient of \ce{CH3 + N -> products} to be 8.6$\times$10$^{-11}$ cm$^{3}$s$^{-1}$, and Marston et al.\cite{Reference442} reported the experimental branching ratios to be

\begin{multline*}
\ce{CH3 + N -> H2CN + H} \\ \Phi \sim 0.9,
\end{multline*}
and
\begin{multline*}
\ce{CH3 + N -> HCN + H2} \\ \Phi \sim 0.1.
\end{multline*}

However, we find only the first of these reactions has an efficient rate coefficient (k = 3.3$\times$10$^{-11}$ cm$^{3}$s$^{-1}$), and that the second reaction is very inefficient (k $\sim$ 10$^{-28}$ cm$^{3}$s$^{-1}$). This result agrees with past theoretical work, which suggests the measurement of the second reaction likely corresponds to a series of reactions passing through the H$_2$CN intermediate \cite{Reference464}. For more details of our analysis, see theoretical case study 4 in SI.

%We find three spin configurations for the reaction \ce{CH3 + N -> H2CN + H}, all of which we find have efficient rate coefficients. These spin configurations are \ce{CH3 + ^4N -> H2CN + H} on the triplet surface, and \ce{CH3 + ^2N -> H2CN + H} on the singlet and triplet surfaces (for mechanistic models, see Figure~\ref{NCH3-model} in Appendix D). We calculate the rate coefficients to be 3.3$\times$10$^{-11}$, 3.1$\times$10$^{-11}$, and 1.0$\times$10$^{-10}$ cm$^{3}$s$^{-1}$, respectively.

%There is one spin configuration for \ce{CH3 + N -> HCN + H2}, corresponding to \ce{CH3 + ^2N -> HCN + H2} on the singlet surface. Contrary to experiment, we find this reaction to be very inefficient, with a rate coefficient of $\sim$10$^{-28}$ cm$^{3}$s$^{-1}$. Moreover, we find no direct route at all for the reported reaction \ce{CH3 + N -> HCN + 2H}.

%\citet{Reference442} suggested the direct pathway \ce{CH3 + N -> HCN + H2} was the simplest explanation for the hydrogen molecules measured in their experiments, however they suggest the reaction of H$_2$CN may be an alternate source that cannot be ruled out. We agree with this latter explanation, as our theoretical work suggests the most likely reaction pathways to HCN from N and CH$_3$ occur through the H$_2$CN intermediate, i.e. \ce{H2CN + H -> HCN + H2}, \ce{H2CN -> HCN + H}, \ce{H2CN + N -> HCN + NH}. These alternate pathways can account for the hydrogen atoms attributed to the reaction \ce{CH3 + N -> HCN + 2H}.

Our theoretical reaction rate coefficients are within an order of magnitude of the closest experimental or suggested value from the literature 93$\%$ of the time. The theoretical reaction rate coefficients for \ce{H2CN + ^4N -> HCN + ^3NH}, \ce{2CH3 -> C2H6}, and \ce{2H2CN -> HCN + H2CNH}, on the other hand, differ by factors of 47, 48, and 89 from the closest experimental values, respectively. In the case of \ce{2H2CN -> HCN + H2CNH}, we are unable to converge the calculations beyond a H-N bond distance of 1.95 $\AA$, and in this case, the rate coefficients increase towards the experimental values with increasing H-N bond distance. Therefore we expect the major source of discrepancy between theory and experiment for this reaction is due to computational convergence. With regards to the other two reactions, we find the discrepancies to be due to our chosen computational method. Calculations at the CCSD/aug-cc-pVDZ level of theory bring the rate coefficient for \ce{2CH3 -> C2H6} to within its experimental range. CCSD calculations, however, do not universally increase accuracy. The rate coefficient for \ce{H2CN + ^4N -> HCN + ^3NH} when calculated using CCSD/aug-cc-pVDZ is over 3 orders of magnitude smaller than the experimental value. On the other hand, this reaction rate coefficient comes to within $\sim$80\% of the experimental value when using CAM-B3LYP/aug-cc-pVDZ. Because CAM-B3LYP has less short-range HF exchange than BHandHLYP\cite{Reference597}, this method is expected to predict a smaller barrier height than BHandHLYP. Thus in this case, where BHandHLYP overestimates the barrier height (underestimates the rate coefficient) with respect to the experimental value, CAM-B3LYP brings the calculated rate coefficient closer to the experimental value. Of future interest would be to test the accuracy of all the rate coefficients in our network when calculated with CAM-B3LYP/aug-cc-pVDZ.

\subsection*{Temperature Dependencies}

\begin{figure}[hbtp]
\centering
\includegraphics[width=\linewidth]{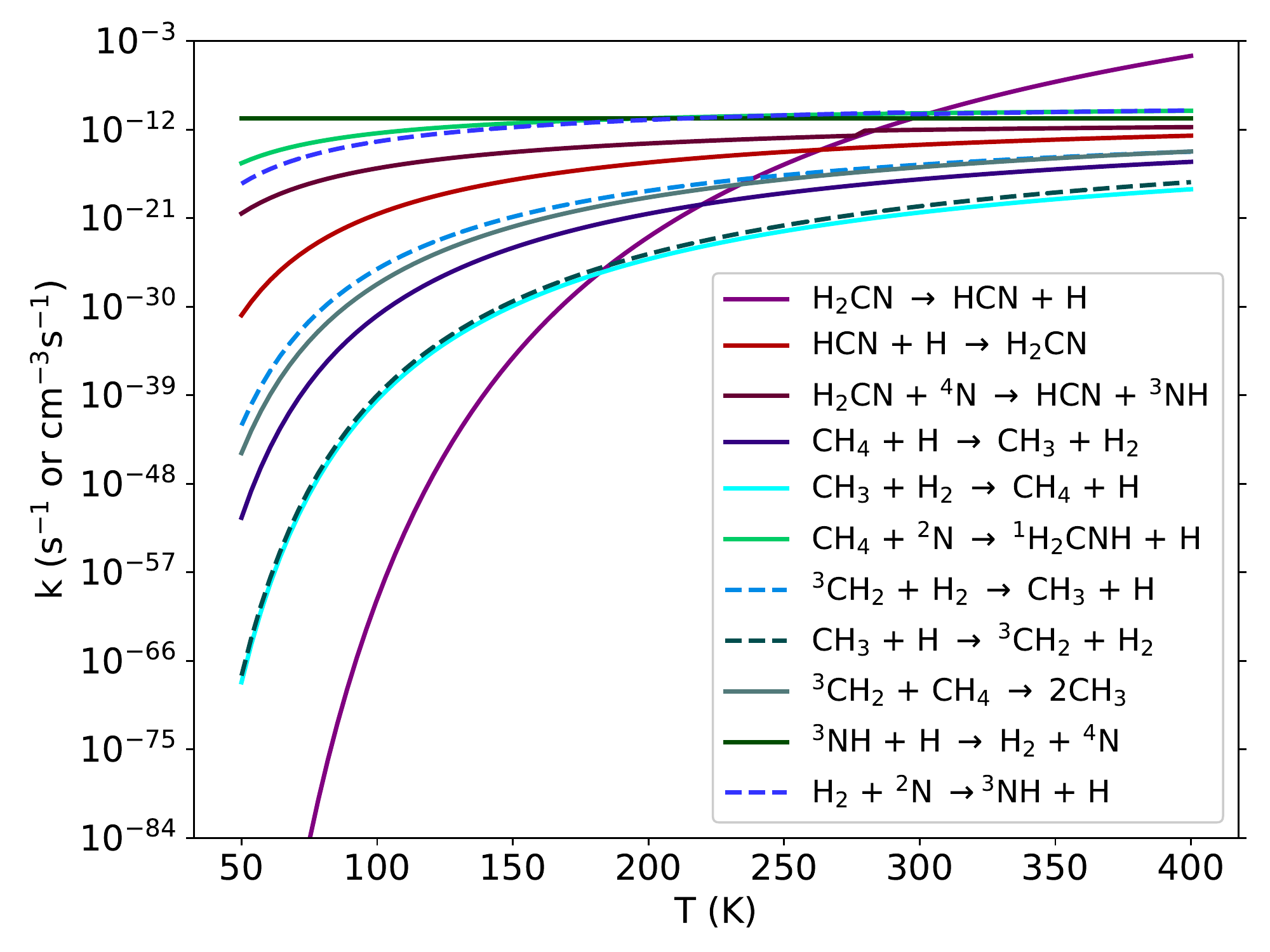}
\caption{Temperature dependence of the 11 reactions in our network that have barriers. Rate coefficients are calculated at 50, 100, 200, 298, and 400 K, and are fit to the Arhennius expression $k(T) = \alpha \left(\frac{T}{300}\right)^{\beta} e^{-\gamma/T}$. Two of the fits have discontinuities: \ce{H2CN + ^4N -> HCN + ^3NH} at 279 K, and \ce{H2 + ^2N -> ^3NH + H} at 304 K. First-order rate coefficients have units s$^{-1}$. Second-order rate coefficients have units cm$^{3}$s$^{-1}$.}
\label{Figure2}
\end{figure}

In Table~\ref{TableTempDep}, we display the Arrhenius coefficients for the reactions in this network for temperatures between 50 and 400 K. We also display the temperature-dependent rate coefficients for the 10 reactions that have barriers in Figure~\ref{Figure2}.

\setlength\LTcapwidth{\textwidth}
\begin{longtable*}{cccccc}
\caption{Arrhenius coefficients for the 42 reactions in this network. Rate coefficients are calculated for the reactions with barriers at 50, 100, 200, 298, and 400 K, and are fit to the Arhennius expression $k(T) = \alpha \left(\frac{T}{300}\right)^{\beta} e^{-\gamma/T}$. Barrierless reaction rate coefficients typically do not vary by more than a factor of 1--3 for temperatures between 50 and 400 K \cite{Reference587,Reference495,Reference481,Reference577,Reference572}, therefore for feasibility of calculations we set the $\beta$ and $\gamma$ for these reactions to zero. For the majority of reactions, fits to the Arrhenius expression are continuous in the temperature range from 50--400 K; however, for two reactions there are discontinuities and thus these reactions have two sets of Arrhenius coefficients.\label{TableTempDep}} \\
Reaction equation & Forward or Reverse? & Temperature range (K) & $\alpha$ &  $\beta$ & $\gamma$ \\ \hline \\[-2mm]
\ce{H2CN <-> HCN + H} & F & 50--400 & 7.9$\times$10$^{13}$ & 0 & 16952\\
 & R & 50--400 & 6.5$\times$10$^{-11}$ & 0.7 & 2318\\
\ce{H2CN + H -> HCN + H2} & F & 50--400 & 1.8$\times$10$^{-11}$ & 0 & 0\\
\ce{H2CN + ^4N <-> HCN + ^3NH} & F & 50--279 & 7.8$\times$10$^{-12}$ & 1.63 & 938 \\
 & F & 279--400 & 1.2$\times$10$^{-11}$ & 0 & 758 \\
\ce{2H2CN <-> HCN + H2CNH} & F & 50--400 & 3.7$\times$10$^{-14}$ & 0 & 0 \\
\ce{CH4 + H <-> CH3 + H2} & F & 50--400 & 5.5$\times$10$^{-11}$ & 0.6 & 4689\\
 & R & 50--400 & 1.5$\times$10$^{-11}$ & -0.32 & 6632\\
\ce{CH4 + ^2N <-> H3CNH <-> ^1H2CNH + H} & F & 50--400 & 4.7$\times$10$^{-10}$ & 0 & 700\\
\ce{CH3 + H <-> CH4} & F & 50--400 & 7.9$\times$10$^{-11}$ & 0 & 0 \\
\ce{CH3 + ^4N <-> ^3H3CN* <-> H2CN + H} & F & 50--400 & 3.3$\times$10$^{-11}$ & 0 & 0 \\
\ce{CH3 + ^2N <-> ^3H3CN* <-> H2CN + H} & F & 50--400 & 1.0$\times$10$^{-10}$ & 0 & 0 \\
\ce{CH3 + ^2N <-> ^1H3CN* <-> H2CN + H} & F & 50--400 & 3.1$\times$10$^{-11}$ & 0 & 0 \\
\ce{2CH3 <-> C2H6} & F & 50--400 & 7.3$\times$10$^{-13}$ & 0 & 0\\
\ce{^1CH2 + H <-> CH3} & F & 50--400 & 8.4$\times$10$^{-11}$ & 0 & 0 \\
\ce{^1CH2 + H2 <-> CH4} & F & 50--400 & 1.0$\times$10$^{-11}$ & 0 & 0 \\
\ce{^1CH2 + ^4N <-> ^4H2CN} & F & 50--400 & 1.1$\times$10$^{-10}$ & 0 & 0 \\
\ce{^1CH2 + ^2N <-> ^2H2CN} & F & 50--400 & 1.5$\times$10$^{-10}$ & 0 & 0\\
\ce{^1CH2 + ^1CH2 <-> C2H4} & F & 50--400 & 9.9$\times$10$^{-12}$ & 0 & 0 \\
\ce{^1CH2 + ^3CH2 <-> C2H4} & F & 50--400 & 3.5$\times$10$^{-11}$ & 0 & 0\\
\ce{^1CH2 + CH3 <-> C2H5* <-> C2H4 + H} & F & 50--400 & 2.3$\times$10$^{-11}$ & 0 & 0 \\
\ce{^1CH2 + CH4 <-> C2H6} & F & 50--400 & 6.1$\times$10$^{-13}$ & 0 & 0\\
\ce{^3CH2 + H <-> CH3} & F & 50--400 & 5.6$\times$10$^{-10}$ & 0 & 0 \\
\ce{^3CH2 + H2 <-> CH3 + H} & F & 50--400 & 5.4$\times$10$^{-11}$ & 0 & 3661 \\
 & R & 50--400 & 4.2$\times$10$^{-11}$ & 0.82 & 6504 \\
\ce{^3CH2 + ^4N <-> ^2H2CN} & F & 50--400 & 1.3$\times$10$^{-10}$ & 0 & 0\\
\ce{^3CH2 + ^2N <-> ^2H2CN} & F & 50--400 & 2.7$\times$10$^{-10}$ & 0 & 0\\
\ce{^3CH2 + ^2N <-> ^4H2CN} & F & 50--400 & 4.3$\times$10$^{-10}$ & 0 & 0\\
\ce{^3CH2 + ^3CH2 <-> C2H4} & F & 50--400 & 4.2$\times$10$^{-11}$ & 0 & 0 \\
\ce{^3CH2 + CH3 <-> C2H5* <-> C2H4 + H} & F & 50--400 & 8.8$\times$10$^{-12}$ & 0 & 0 \\
\ce{^3CH2 + CH4 <-> 2CH3} & F & 50--400 & 5.5$\times$10$^{-11}$ & 1.63 & 3840 \\
 & R & 50--400 & 5.5$\times$10$^{-11}$ & 0 & 0\\
\ce{CH + H <-> ^1CH2} & F & 50--400 & 1.5$\times$10$^{-10}$ & 0 & 0 \\
\ce{CH + H <-> ^3CH2} & F & 50--400 & 5.3$\times$10$^{-10}$ & 0 & 0\\
\ce{CH + H2 <-> CH3} & F & 50--400 & 7.9$\times$10$^{-11}$ & 0 & 0 \\
\ce{CH + ^4N <-> ^3HCN <-> CN + H} & F & 50--400 & 1.1$\times$10$^{-10}$ & 0 & 0\\
\ce{CH + ^2N <-> ^3HCN <-> CN + H} & F & 50--400 & 2.7$\times$10$^{-10}$ & 0 & 0\\
\ce{2CH <-> C2H2} & F & 50--400 & 1.3$\times$10$^{-11}$ & 0 & 0\\
\ce{CH + CH4 <-> C2H5* <-> C2H4 + H} & F & 50--400 & 3.8$\times$10$^{-13}$ & 0 & 0\\
\ce{^3NH + H <-> H2 + ^4N} & F & 50--400 & 1.4$\times$10$^{-11}$ & 0 & 0\\
\ce{^3NH + H <-> NH2* <-> H2 + ^2N} & R & 50--304 & 1.1$\times$10$^{-9}$ & 0.83 & 909 \\
& R & 304--400 & 1.5$\times$10$^{-9}$ & 0 & 1128 \\
\ce{^3NH + ^4N <-> N2H* <-> N2 + H} & F & 50--400 & 4.0$\times$10$^{-11}$ & 0 & 0 \\
\ce{^3NH + ^2N <-> N2H* <-> N2 + H} & F & 50--400 & 5.5$\times$10$^{-11}$ & 0 & 0\\ 
\hline
\end{longtable*}

The majority of the reactions with barriers fit to one Arrhenius expression for the 50--400 K temperature range, however there were two special cases that had discontinuous fits. Both \ce{H2CN + ^4N -> HCN + ^3NH} and \ce{H2 + ^2N -> NH2* -> ^3NH + H} have two Gibbs maxima along their MEP's. As temperature increases, the shorter of the Gibbs humps increases in height until it reaches the same height as the other hump at some characteristic temperature. Beyond this temperature, the previously shorter Gibbs hump surpasses the other hump in height, becoming the new location of the variational transition state. Such a drastic change in the location of the variational transition state before and after the characteristic temperature creates a discontinuity in the temperature dependent rate coefficient, that is better fit to two separate sets of Arrhenius coefficients.

The rate coefficients of four of the reactions with barriers do not decrease rapidly with decreasing temperatures, and remain ``fast'' (k $>$ 10$^{-21}$ cm$^{3}$s$^{-1}$) in the entire 50--400 K temperature range:

\begin{equation*}
\ce{CH4 + ^2N -> H3CNH -> ^1H2CNH + H},
\end{equation*}
\begin{equation*}
\ce{^3NH + H -> H2 + ^4N},
\end{equation*}
\begin{equation*}
\ce{H2 + ^2N -> NH2* -> ^3NH + H},
\end{equation*}
\begin{equation*}
\ce{H2CN + ^4N -> HCN + ^3NH}.
\end{equation*}

The rate coefficients of the other seven reactions with barriers drop off more rapidly for colder temperatures, and become ``slow'' in the $\sim$100--300 K range. One reaction's rate coefficient has a particularly interesting temperature dependence. \ce{H2CN -> HCN + H} has a rate coefficient as high as 3.1$\times$10$^{-5}$ s$^{-1}$ at 400 K, and as low as 4.5$\times$10$^{-134}$ s$^{-1}$ at 50 K. Above 320 K, \ce{H2CN -> HCN + H} has the highest rate coefficient in this network.

\subsection*{Effects of Spin Configuration on Rate Coefficients}

Both ground state (e.g. $^4$N, $^3$CH$_2$) and excited state (e.g. $^2$N, $^1$CH$_2$) species are produced during the UV photodissociation of N$_2$ and CH$_4$. Because our network includes both ground state and excited state species, there is often more than one possible spin configuration for a given reaction. For example, the reaction 

\begin{equation*}
\ce{CH3 + N -> H2CN + H}
\end{equation*}
has three spin configurations. If the nitrogen is in the ground state, the reaction passes through the excited state $^3$H$_3$CN intermediate before decaying into H$_2$CN + H directly, or after passing though the $^3$H$_2$CNH intermediate. If the nitrogen is in the excited state, the reaction can either pass through the excited state $^3$H$_3$CN intermediate, or the ground state $^1$H$_3$CN intermediate, before decaying into H$_2$CN + H directly, or after passing through the $^3$H$_2$CNH or $^1$H$_2$CNH intermediates. In other words, on the triplet PES there are two possible reactions: \ce{CH3 + ^4N -> ^3H3CN -> H2CN + H} and \ce{CH3 + ^2N -> ^3H3CN -> H2CN + H}, and on the singlet PES there is one reaction: \ce{CH3 + ^2N -> ^1H3CN -> H2CN + H}. The first steps of these reactions are the rate-limiting steps, and these steps are barrierless. All reactions have the same products, a ground state H$_2$CN molecule and H atom. However, the rate coefficient for \ce{CH3 + ^2N -> ^3H3CN -> H2CN + H} is larger than the other two reactions by a factor of 3 (see Table~\ref{Table3} for calculated values).

Rate coefficients for different reaction spin configurations can also vary by several orders of magnitude, especially if a reaction barrier exists. The reaction \ce{H2CN + N -> HCN + NH} has three spin configurations that produce ground state HCN. On the singlet surface, there is \ce{H2CN + ^2N -> HCN + ^1NH}, and on the triplet surface, there is \ce{H2CN + ^4N -> HCN + ^3NH} and \ce{H2CN + ^2N -> HCN + ^3NH}. All these reactions have an energy barrier, but only the spin configuration involving the $^4$N atom is efficient. We calculate the rate coefficient for \ce{H2CN + ^4N -> HCN + ^3NH} to be 9.4$\times$10$^{-13}$ cm$^{3}$s$^{-1}$, which is 16 and 18 orders of magnitude larger than our calculated rate coefficients for \ce{H2CN + ^2N -> HCN + ^3NH} and \ce{H2CN + ^2N -> HCN + ^1NH}, respectively.

Different spin configurations for two reactants can also lead to different products. For example, when $^1$CH$_2$ and CH$_4$ react on the singlet surface, they come together to form C$_2$H$_6$. When the hydrogen from CH$_4$ bonds with the carbon of $^1$CH$_2$, the resultant CH$_3$ molecules each have an unpaired electron of opposite spin, allowing these molecules to rapidly bond to form C$_2$H$_6$. However, when $^3$CH$_2$ and CH$_4$ react on the triplet surface, they react directly to form two CH$_3$ molecules, each with an unpaired electron of the same spin. The rate coefficient of \ce{^1CH2 + CH4 -> C2H6} is also 5 orders of magnitude larger than \ce{^3CH2 + CH4 -> 2CH3}. This is largely due to the fact that \ce{^1CH2 + CH4 -> C2H6} is barrierless, whereas \ce{^3CH2 + CH4 -> 2CH3} has an energy barrier.

\section*{Conclusions}\label{conclusions}

In this work, we use canonical variational transition state theory (CVT) to calculate 42 rate coefficients that are directly involved with or are in competition with HCN production in early Earth or Titan atmospheres. Approximately 36\% of these reactions have no previously reported experimental or suggested value. To make such a large network of calculations feasible, we make use of computational quantum chemistry simulations at an accurate yet inexpensive level of theory: BHandHLYP/aug-cc-pVDZ. Moreover, we only calculate the temperature dependence of the rate coefficients for the reactions that have barriers. By using one level of theory for all reaction rate coefficient calculations, we expect the computational errors to be similar. 

In this network, we focus on HCN production from methane and nitrogen radicals, which are produced in the atmosphere via UV photodissociation or lightning. Dissociation of CH$_4$ and N$_2$ produces both excited and ground state species, therefore we calculate the rate coefficients for multiple spin configurations involving these species. The reactions in our network have 1--5 spin configurations. 

We list our five most important results below.

\begin{itemize}
\item We provide consistently calculated rate coefficients for 15 reactions that have no previously suggested values. In this sense, we fill a substantial gap in the data. These previously unknown rate coefficients include those of several key reactions in the pathway to produce atmospheric HCN (e.g. \ce{CH2 + N -> H2CN} and \ce{H2CN -> HCN + H} \cite{2007AsNow..22e..76R}).
\item Of the reactions in our network with past experimental or suggested values, 93\% are within an order of magnitude of these values. The remaining 7\% differ by less than 2 orders of magnitude from experimental values. These discrepancies are either due to convergence issues or our chosen computation method. When convergence isn't an issue, re-running rate coefficient calculations at the similarly expensive CAM-B3LYP/aug-cc-pVDZ level of theory or the more expensive CCSD/aug-cc-pVDZ level of theory decreases the discrepancy between theory and experimental values.
\item We find the reaction of \ce{CH3 + N -> HCN + H2} on the singlet surface to be inefficient, with a rate coefficient near 10$^{-28}$ cm$^{3}$s$^{-1}$ (confirming the results of Cimas and Largo\cite{Reference464}). This is in contrast to experimental results which suggest a rate coefficient to have a value near 10$^{-11}$ cm$^{3}$s$^{-1}$ \cite{Reference442}. The experimental result may be due to the measurement of multi-step reaction, e.g., \ce{CH3 + N -> H2CN + H} and \ce{H2CN + H -> HCN + H2}. However, we cannot exclude the possibility of a spin-forbidden process accounting for this experimental value.
\item The effects of reaction spin configuration on the rate coefficient can be both subtle and substantial. For a given reaction, differences in rate coefficients between spin configurations can range from factors of order unity, up to 18 orders of magnitude. If there is a barrier involved with one or more of the reaction spin configurations, the difference between their reaction rate coefficients tends to be much greater than if all the reaction spin configurations are barrierless.
\item Seven reaction rate coefficients in our network decrease rapidly with decreasing temperature, and become ``slow'' ($k <$ 10$^{-21}$) at temperatures below $\sim$100--300 K. One reaction, \ce{H2CN -> HCN + H}, increases rapidly for increasing temperatures; above 320 K, this reaction has the highest rate coefficient in the network.
\end{itemize}

Overall, we find CVT and computational quantum chemistry simulations at the BHandHLYP/aug-cc-pVDZ level of theory to be a feasible and accurate method for calculating a large set of small-molecule, multiple-spin configuration reaction rate coefficients for a range of terrestrial atmospheric temperatures. We also note that although calculations at the CCSD/aug-cc-pVDZ level of theory often lead to improvements in the rate coefficients' conformance to experimental values, computational cost and convergence issues made calculating all the rate coefficients at this level of theory impossible. Based on a limited number of calculations, we also find CAM-B3LYP to be an accurate alternative functional for performing CVT rate coefficient calculations and recommend it for a wider study.

\section*{Acknowledgments}

We thank the two anonymous referees, whose reports led to the improvement of this work. We gratefully acknowledge Anand Patel (McMaster) and Farnaz Heidar-Zadeh (Univ. of Luxembourg) whose previous work and comments helped in the development of this research. B.K.D.P. is supported by an NSERC Postgraduate Scholarship-Doctoral (PGS-D). P.W.A is supported by NSERC, the Canada Research Chairs, and Canarie. R.E.P. is supported by an NSERC Discovery Grant. We are grateful to Compute Canada for the computer time allocated for this research.

\section*{References}

\bibliography{Bibliography_Rates}

\beginsupplement

\section*{Supporting Information}

\subsection*{Experimental Data}

Experiments and reviews have measured and suggested reaction rate coefficients for several of the reactions in this network at or near $\sim$298 K. These values are listed in Table~\ref{TableExp}.

\setlength\LTcapwidth{\textwidth}
\begin{longtable*}{ccccc}
\caption{All available experimental or recommended reaction rate coefficients for the reactions in this study. For brevity, only the 13 most recent experimental rate coefficients are listed for \ce{CH3 + CH3 -> C2H6}, for a complete listing, we refer the reader to the NIST Chemical Kinetics Database\cite{Reference494}. First-order rate coefficients have units s$^{-1}$. Second-order rate coefficients have units cm$^{3}$s$^{-1}$.\label{TableExp}} \\
k(298K) & Technique & Temp. (K) & Pressure (Torr) & Reference(s)\\ \hline \\[-2mm]
\ce{H2CN + H -> HCN + H2} & & & & \\
8.3$\times$10$^{-11}$ & Z & independent & &	Tomeczek and Grado{\'n}\cite{Reference446}\\
\ce{H2CN + ^4N -> HCN + ^3NH} & & & & \\
4.4$\times$10$^{-11}$ & M & 298 & 1 & Nesbitt et al.\cite{Reference573}\\
\ce{2H2CN -> HCN + H2CNH} & & & & \\
3.3--8.3$\times$10$^{-12}$ & M & 300 & 120--480 &	Horne and Norrish\cite{Reference447}\\
\ce{CH4 + H -> CH3 + H2} & & & & \\
3.5$\times$10$^{-17}$ & M & 298 &  & Lawrence and Firestone\cite{Reference452}\\
1.7$\times$10$^{-17}$ & M & 298 & 0.55 & Jones and Ma\cite{Reference450}\\
8.2$\times$10$^{-19}$ & S & 300 & & Baulch et al.\cite{Reference451}\\
\ce{CH4 + ^2N -> H2CNH + H} & & & & \\
$^a$4.5$\times$10$^{-12}$ & M & 298 & 700 & Takayanagi et al.\cite{Reference579},\\
& & & & Umemoto et al.\cite{Reference580}\\
$^a$3.7$\times$10$^{-12}$ & M & 300 & 6 & Fell et al.\cite{Reference555},\\
& & & & Umemoto et al.\cite{Reference580}\\
$^a$2.7$\times$10$^{-12}$ & M & 295 & 20 & Umemoto et al.\cite{Reference581},\\
& & & & Umemoto et al.\cite{Reference580}\\
$^a$2.4$\times$10$^{-12}$ & M & 300 & 3--5 & Black et al.\cite{Reference559},\\
& & & & Umemoto et al.\cite{Reference580}\\
$^a$3.2$\times$10$^{-12}$ & S & 298 & & Herron\cite{Reference549},\\
& & & & Umemoto et al.\cite{Reference580}\\
\ce{CH4 + ^2N -> CH3 + ^3NH} & & & & \\
$^b$1.7$\times$10$^{-12}$ & M & 298 & 700 & Takayanagi et al.\cite{Reference579},\\
& & & & Umemoto et al. \cite{Reference580}\\
$^b$1.4$\times$10$^{-12}$ & M & 300 & 6 & Fell et al.\cite{Reference555},\\
& & & & Umemoto et al.\cite{Reference580}\\
$^b$1.0$\times$10$^{-12}$ & M & 295 & 20 & Umemoto et al.\cite{Reference581},\\
& & & & Umemoto et al.\cite{Reference580}\\
$^b$9.0$\times$10$^{-13}$ & M & 300 & 3--5 & Black et al.\cite{Reference559},\\
& & & & Umemoto et al.\cite{Reference580}\\
$^b$1.2$\times$10$^{-12}$ & S & 298 & & Herron\cite{Reference549},\\
& & & & Umemoto et al.\cite{Reference580}\\
\ce{CH3 + H -> CH4} & & & & \\
4.7$\times$10$^{-10}$ & M & 300 & high-pressure limit & Brouard et al.\cite{Reference468}\\
3.3$\times$10$^{-10}$ & M & 308 & high-pressure limit & Cheng and Yeh\cite{Reference470}\\
2.5$\times$10$^{-10}$ & M & 308 & 300 & Cheng et al.\cite{Reference471}\\
2.0$\times$10$^{-10}$ & M & 296 & 735--755 & Sworski et al.\cite{Reference473}\\
1.5$\times$10$^{-10}$ & M & 300 & high-pressure limit & Patrick et al.\cite{Reference474}\\
3.4$\times$10$^{-10}$ & F & 298 & high-pressure limit & Michael et al.\cite{Reference475}\\
3.5$\times$10$^{-10}$ & S & independent & high-pressure limit & Cobos and Troe\cite{Reference469}\\
3.5$\times$10$^{-10}$ & S & independent & high-pressure limit & Baulch et al.\cite{Reference451}\\
2.0$\times$10$^{-10}$ & S &  298 & high-pressure limit & Tsang\cite{Reference472}\\
\ce{CH3 + H2 -> CH4 + H} & & & & \\
$^c$1.3$\times$10$^{-20}$ & M & 300 &  & Kobrinsky and Pacey\cite{Reference588}\\
1.2$\times$10$^{-20}$ & S & 300 &  & Tsang and Hampson\cite{Reference509}\\
9.6$\times$10$^{-21}$ & S & 300 &  & Baulch et al.\cite{Reference451}\\
\ce{CH3 + ^4N -> HCN + H2} & & & & \\
8.6$\times$10$^{-12}$ & M & 298 & 0.3--1.6 & Marston et al.\cite{Reference442},\\
& & & & Stief et al.\cite{Reference444}\\
\ce{CH3 + ^4N -> H2CN + H} & & & & \\
7.7$\times$10$^{-11}$ & M & 298 & 0.3--1.6 &	Marston et al.\cite{Reference442},\\
& & & & Stief et al.\cite{Reference444}\\
5.0$\times$10$^{-11}$ & S & independent & & Miller and Bowman\cite{Reference445}\\
\ce{2CH3 -> C2H6} & & & & \\
6.5$\times$10$^{-11}$ & M & 300 & high-pressure limit & Walter et al.\cite{Reference484}\\
6.5$\times$10$^{-11}$ & M & 298 & high-pressure limit & Macpherson et al.\cite{Reference493}\\
6.0$\times$10$^{-11}$ & M & 298 & high-pressure limit & Du et al.\cite{Reference483}\\
6.0$\times$10$^{-11}$ & M & 298 & high-pressure limit & Slagle et al.\cite{Reference489}\\
$^d$5.9$\times$10$^{-11}$ & M & 292 & 758 & Sangwan et al.\cite{Reference486}\\
5.8$\times$10$^{-11}$ & M & 298 & 750 & Pagsberg et al.\cite{Reference490}\\
5.5$\times$10$^{-11}$ & M & 298 & high-pressure limit & Hippler et al.\cite{Reference487}\\
5.2$\times$10$^{-11}$ & M & 298 & 100 & Fahr et al.\cite{Reference485}\\
4.6$\times$10$^{-11}$ & M & 300 & 1 & Wang et al.\cite{Reference513}\\
4.0$\times$10$^{-11}$ & M & 302 & 81--571 & Arthur\cite{Reference492}\\
3.5$\times$10$^{-11}$ & M & 308 & 86 & Anastasi and Arthur\cite{Reference491}\\
6.0$\times$10$^{-11}$ & S & independent & high-pressure limit & Baulch et al.\cite{Reference451}\\
4.4$\times$10$^{-11}$ & S & 298 & high-pressure limit & Tsang\cite{Reference472}\\
\ce{^1CH2 + H -> CH + H2} & & & & \\
5.0$\times$10$^{-11}$ & S & & & Tsang and Hampson\cite{Reference509}\\
\ce{^1CH2 + H2 -> CH4* -> CH3 + H} & & & & \\
1.3$\times$10$^{-10}$ & M & 295 & 6 & Langford et al.\cite{Reference514}\\
1.1$\times$10$^{-10}$ & M & 298 & 10$^{-4}$--10 & Ashfold et al.\cite{Reference515}\\
7.0$\times$10$^{-12}$ & M & 298 & 10 & Braun et al.\cite{Reference510}\\
1.2$\times$10$^{-10}$ & S &  & & Tsang and Hampson\cite{Reference509}\\
1.2$\times$10$^{-10}$ & S & independent & & Baulch et al.\cite{Reference451}\\
\ce{^1CH2 + ^1CH2 -> C2H2 + 2H} & & & & \\
5.0$\times$10$^{-11}$ & S & & & Tsang and Hampson\cite{Reference509}\\
\ce{^1CH2 + ^3CH2 -> C2H2 + H2} & & & & \\
3.0$\times$10$^{-11}$ & S & & & Tsang and Hampson\cite{Reference509}\\
\ce{^1CH2 + CH3 -> C2H4 + H} & & & & \\
3.0$\times$10$^{-11}$ & S & & & Tsang and Hampson\cite{Reference509}\\
\ce{^1CH2 + CH4 -> 2CH3} & & & & \\
7.3$\times$10$^{-11}$ & M & 298 & 10$^{-4}$--10 & Ashfold et al.\cite{Reference515}\\
7.0$\times$10$^{-11}$ & M & 295 & 6 & Langford et al.\cite{Reference514}\\
1.9$\times$10$^{-12}$ & M & 298 & 5--20 & Braun et al.\cite{Reference510}\\
7.1$\times$10$^{-11}$ & S & & & Tsang and Hampson\cite{Reference509}\\
\ce{^3CH2 + H -> CH3* -> CH + H2} & & & & \\
2.7$\times$10$^{-10}$ & M & 285 & 2 & Boullart and Peeters\cite{Reference504}\\
2.7$\times$10$^{-10}$ & M & 298 & 2 & B{\"o}hland and Temps\cite{Reference507}\\
$^e$2.6$\times$10$^{-10}$ & M & 300 & 2 & Devriendt et al.\cite{Reference503}\\
1.8$\times$10$^{-10}$ & M & 298 & 1--2 & B{\"o}hland et al.\cite{Reference505}\\
$^c$1.4$\times$10$^{-10}$ & M & 300 & 100 & Zabarnick et al.\cite{Reference506}\\
8.3$\times$10$^{-11}$ & M & 298 & 2 & Grebe and Homann\cite{Reference508}\\
2.0$\times$10$^{-10}$ & S & 300 & & Baulch et al.\cite{Reference451}\\
2.7$\times$10$^{-10}$ & S & 298 & & Tsang and Hampson\cite{Reference509}\\
\ce{^3CH2 + H2 -> CH3 + H} & & & & \\
$<$5.0$\times$10$^{-14}$ & M & 298 & 10 & Braun et al.\cite{Reference510}\\
$<$6.9$\times$10$^{-15}$ & M & 295 & 8 & Darwin and Moore\cite{Reference545}\\
$<$5.0$\times$10$^{-15}$ & M & & 10 & Pilling and Robertson\cite{Reference544}\\
\ce{^3CH2 + ^3CH2 -> C2H2 + 2H} & & & & \\
5.3$\times$10$^{-11}$ & M & 298 & 20--700 & Braun et al.\cite{Reference510}\\
5.3$\times$10$^{-11}$ & S & 300 & & Baulch et al.\cite{Reference451}\\
\ce{^3CH2 + CH3 -> C2H5* -> C2H4 + H} & & & & \\
2.1$\times$10$^{-10}$ & M & 300 & 1 & Wang and Fockenberg\cite{Reference513}\\
1.1$\times$10$^{-10}$ & M & 298 & 1 & Deters et al.\cite{Reference578}\\
1.0$\times$10$^{-10}$ & M & 308 & 50--700 & Laufer and Bass\cite{Reference512}\\
5.0$\times$10$^{-11}$ & M & & 200 & Pilling and Robertson\cite{Reference511}\\
7.0$\times$10$^{-11}$ & S & independent & & Baulch et al.\cite{Reference451}\\
7.0$\times$10$^{-11}$ & S & 298 & & Tsang and Hampson\cite{Reference509}\\
\ce{^3CH2 + CH4 -> 2CH3} & & & & \\
$<$5.0$\times$10$^{-14}$ & M & 298 & 10 & Braun et al.\cite{Reference510}\\
$^f$3.1$\times$10$^{-19}$ & M & 298 & 2--3 & B{\"o}hland et al.\cite{Reference568}\\
$<$3.0$\times$10$^{-19}$ & S & 298 & & Tsang and Hampson\cite{Reference509}\\
\ce{CH + H2 -> CH3} & & & & \\
1.6$\times$10$^{-10}$ & M & 294 & high-pressure limit & Brownsword et al.\cite{Reference575}\\
5.1$\times$10$^{-11}$ & M & 300 & 750 & Fulle and Hippler\cite{Reference516}\\
4.5$\times$10$^{-11}$ & M & 298 & 591 & Becker et al.\cite{Reference517}\\
4.5$\times$10$^{-11}$ & M & 279 & 600 & Berman and Lin\cite{Reference518}\\
3.0$\times$10$^{-11}$ & M & 294 & 750 & McIlroy and Tully\cite{Reference576}\\
2.3$\times$10$^{-11}$ & M & 298 & 100 & Butler et al.\cite{Reference521}\\
1.7$\times$10$^{-11}$ & M & & & Bosnali and Perner\cite{Reference520}\\
1.0$\times$10$^{-12}$ & M & 298 & 1--9 & Braun et al.\cite{Reference519}\\
\ce{CH + H2 -> CH3* -> ^3CH2 + H} & & & & \\
$^c$9.1$\times$10$^{-13}$ & M & 300 & 100 & Zabarnick et al.\cite{Reference506}\\
1.2$\times$10$^{-12}$ & M & 294 & 400 & Brownsword et al.\cite{Reference575}\\
4.5$\times$10$^{-11}$ & M & 298 & 591 &  Becker et al.\cite{Reference517}\\
\ce{CH + ^4N -> ^3HCN -> CN + H} & & & & \\
1.6$\times$10$^{-10}$ & M & 298 & 4 & Brownsword et al.\cite{Reference570}\\
1.2--1.4$\times$10$^{-10}$ & M & 296 & 5 & Daranlot et al.\cite{Reference572}\\
2.1$\times$10$^{-11}$ & M & 298 & 5--15 & Messing et al.\cite{Reference571}\\
\ce{2CH -> C2H2} & & & & \\
2.0$\times$10$^{-10}$ & M & 298 & 1--330 & Braun et al.\cite{Reference519}\\
1.7$\times$10$^{-10}$ & M & 298 & 1--500 & Braun et al.\cite{Reference523}\\
\ce{CH + CH4 -> C2H4 + H} & & & & \\
3.0$\times$10$^{-10}$ & M & 298 & 30--100 & Butler et al.\cite{Reference521}\\
1.0$\times$10$^{-10}$ & M & 298 & 100 & Butler et al.\cite{Reference562}\\
9.8$\times$10$^{-11}$ & M & 298 & 100 &  Berman and Lin\cite{Reference561}\\
9.1$\times$10$^{-11}$ & M & 298 & 50--300 & Blitz et al.\cite{Reference563}\\
8.9$\times$10$^{-11}$ & M & 295 & 9--12 & Canosa et al.\cite{1997AA...323..644C}\\
6.7$\times$10$^{-11}$ & M & 298 & 100 & Thiesemann et al.\cite{Reference566}\\
3.3$\times$10$^{-11}$ & M & & & Bosnali and Perner\cite{Reference520}\\
2.5$\times$10$^{-12}$ & M & 298 & 100 & Braun et al.\cite{Reference519}\\
2.0$\times$10$^{-12}$ & M & 298 & 1-500 & Braun et al.\cite{Reference523}\\
9.8$\times$10$^{-11}$ & S & 298 & & Baulch et al.\cite{Reference451}\\
\ce{^3NH + H -> H2 + ^4N} & & & & \\
3.2$\times$10$^{-12}$ & M & 298 & 2--8 & Adam et al.\cite{Reference527}\\
\ce{^3NH + ^4N -> N2 + H} & & & & \\
2.5$\times$10$^{-11}$ & M & 298 & 11--15 & Hack et al.\cite{Reference525}\\
2.6$\times$10$^{-11}$ & S & 300 & & Konnov and De Ruyck\cite{Reference526}\\
\ce{^2N + H2 -> NH2* -> ^3NH + H} & & & & \\
5.0$\times$10$^{-12}$ & M & 300 & 3--5 & Black et al.\cite{Reference559}\\
3.5$\times$10$^{-12}$ & M & 300 & 6 & Fell et al.\cite{Reference555}\\
2.7$\times$10$^{-12}$ & M & 300 & 2--5 & Black et al.\cite{Reference556}\\
2.4$\times$10$^{-12}$ & M & 300 & 753 & Suzuki et al.\cite{Reference552}\\
2.3$\times$10$^{-12}$ & M & 295 & 30 & Umemoto et al.\cite{Reference548}\\
2.3$\times$10$^{-12}$ & M & 300 & 1--3 & Piper et al.\cite{Reference554}\\
2.1$\times$10$^{-12}$ & M & 300 & 26 & Husain et al.\cite{Reference557}\\
1.8$\times$10$^{-12}$ & M & 298 & 1 & Whitefield et al.\cite{Reference553}\\
1.7$\times$10$^{-12}$ & M & 300 & 50 & Husain et al.\cite{Reference558}\\
2.2$\times$10$^{-12}$ & S & 200--300 &  & Herron\cite{Reference549}\\
\hline
\multicolumn{5}{l}{\footnotesize $^a$ Experimental value of \ce{CH4 + ^2N -> products}, multiplied by a branching ratio of 0.8\cite{Reference580}.} \\
\multicolumn{5}{l}{\footnotesize $^b$ Experimental value of \ce{CH4 + ^2N -> products}, multiplied by a branching ratio of 0.3\cite{Reference580}.} \\
\multicolumn{5}{l}{\footnotesize $^c$ Experiments performed at 372 K and extrapolated to 300 K.} \\
\multicolumn{5}{l}{\footnotesize $^d$ Value taken as average of two identical experiments.} \\
\multicolumn{5}{l}{\footnotesize $^e$ Experiments performed at $\geq$ 400 K and extrapolated to 300 K.} \\
\multicolumn{5}{l}{\footnotesize $^f$ Experiments performed at $\geq$ 413 K and extrapolated to 300 K.} \\
\multicolumn{5}{l}{\footnotesize Z: Zero activation energy value. Calculated by numerical modeling using the chemical compositions of the flames of CH$_4$ + air.} \\
\multicolumn{5}{l}{\footnotesize M: Monitoring decay of reactants and/or production of products.} \\
\multicolumn{5}{l}{\footnotesize S: Suggested value based on experiments and/or evaluations at a range of temperatures.} \\
\multicolumn{5}{l}{\footnotesize F: Fitting of simulated concentration profiles to absolute concentration profiles from experiments reported by Barker et al.\cite{Reference476}.} \\
\end{longtable*}

%\ref{Eq12} & \ce{2H2CN -> HCN + CH2NH} & \hspace{3mm}3.3--8.3$\times$10$^{-12}$ & T &	\citet{Reference447}\\

\subsection*{Previous Theoretical Data}

Previous theoretical studies have been performed on the reactions in this network. In Table~\ref{TableTheor}, we list the theoretical rate coefficients and the methods that were employed to calculate them.

\setlength\LTcapwidth{\textwidth}
\begin{longtable*}{cccc}
\caption{Previous theoretical rate coefficients for the reactions in this study. For sources that performed multiple rate coefficient calculations with different theoretical and/or computational methods, we list the range of their results here. First-order rate coefficients have units s$^{-1}$. Second-order rate coefficients have units cm$^{3}$s$^{-1}$.\label{TableTheor}} \\
k(298K) & Theory & Computational Method & Reference(s) \\ \hline \\[-2mm]
\ce{CH4 + H -> CH3 + H2} & & & \\[+1mm]
4.1$\times$10$^{-21}$--2.2$\times$10$^{-18}$ & TST, quantum dynamics &	\hspace{2mm} CCSD(T)/cc-pVTZ$^a$ & Kerkeni and Clary\cite{Reference569}\\
8.4$\times$10$^{-19}$--2.1$\times$10$^{-18}$ & CVT + SCT$^b$ &	\hspace{2mm} BH\&HLYP/6-311G(d,p), & Truong and Duncan\cite{Reference454}\\
 & & \hspace{2mm} PMP4/6-311+G(2df,2pd)$^c$ & \\
1.6$\times$10$^{-18}$ & CVT + SCT &	\hspace{2mm} PMP4/cc-pVTZ$^d$ & Maity et al.\cite{Reference456}\\
1.3$\times$10$^{-18}$ & CVT + SCT$^e$ &	\hspace{2mm} QCISD/6-311G(d,p) & Truong\cite{Reference590}\\
9.8$\times$10$^{-19}$ & CVT + $\mu$OMT &	 & Espinosa-Garc{\'i}a and Corchado\cite{Reference458}, \\
& & & Jordan and Gilbert\cite{Reference459}\\
6.5$\times$10$^{-19}$ & CVT + $\mu$OMT &	 & Espinosa-Garc{\'i}a and Corchado\cite{Reference458}, \\
& & & Joseph et al.\cite{Reference460}\\
3.0$\times$10$^{-21}$--6.0$\times$10$^{-19}$ & TST, quantum dynamics & CCSD(T)/cc-pVTZ$^a$ &  Kerkeni and Clary\cite{Reference455}\\
3.8$\times$10$^{-19}$ & BEBO & & Clark and Dove\cite{Reference462}\\
8.7$\times$10$^{-21}$--2.4$\times$10$^{-19}$ & TST, CVT, CVT + SCT &	\hspace{2mm} & Pu and Truhlar\cite{Reference453}\\
1.8$\times$10$^{-21}$--1.6$\times$10$^{-19}$ & TST &	\hspace{2mm} G2(MP2)$^f$, BAC-MP4$^g$ & Berry et al.\cite{Reference457}\\
8.1$\times$10$^{-20}$ & TST + tunneling & \hspace{2mm} UMP2/6-31G-(d,p) & Bryukov et al.\cite{Reference589}\\
4.6$\times$10$^{-20}$ & TST + WTC &	PMP4SDTQ/6-311G**$^h$ & Gonzalez et al.\cite{Reference461}\\
\ce{CH3 + H2 -> CH4 + H} & & & \\[+1mm]
1.2$\times$10$^{-19}$ & CVT + $\mu$OMT &	 & Espinosa-Garc{\'i}a and Corchado\cite{Reference458}, \\
& & & Joseph et al.\cite{Reference460}\\
1.1$\times$10$^{-19}$ & CVT + SCT &	\hspace{2mm} PMP4/cc-pVTZ$^d$ & Maity et al.\cite{Reference456}\\
1.5--8.2$\times$10$^{-20}$ & CVT + SCT$^b$ &	\hspace{2mm} BH\&HLYP/6-311G(d,p), & Truong and Duncan\cite{Reference454}\\
 & & \hspace{2mm} PMP4/6-311+G(2df,2pd)$^c$ & \\
8.0$\times$10$^{-20}$ & CVT + $\mu$OMT &	 & Espinosa-Garc{\'i}a and Corchado\cite{Reference458}, \\
& & & Jordan and Gilbert\cite{Reference459}\\
5.2$\times$10$^{-20}$ & CVT + SCT$^e$ &	\hspace{2mm} QCISD/6-311G(d,p) & Truong\cite{Reference590}\\
1.7$\times$10$^{-20}$ & TST + tunneling & \hspace{2mm} UMP2/6-31G-(d,p) & Bryukov et al.\cite{Reference589}\\
\ce{CH3 + ^4N -> H3CN* -> H2CN + H} & & & \\[+1mm]
1.9$\times$10$^{-10}$ & CVT + RRKM  & \hspace{2mm} CCSD(T)/CBS & Alves et al.\cite{Reference463}\\
9.1$\times$10$^{-12}$ & $\mu$VT + RRKM  & \hspace{2mm} CCSD(T)/cc-pVTZ$^i$ & Cimas and Largo\cite{Reference464}\\
\ce{CH4 + ^2N -> products} & & & \\
8.5$\times$10$^{-14}$ & CVT & \hspace{2mm} CASSCF(5,5)/6-311G** & Takayanagi et al.\cite{Reference579}\\
\ce{CH3 + H -> CH4} & & & \\[+1mm]
6.7$\times$10$^{-10}$ & CVT &	\hspace{2mm} MP4/6-31G** & Hase and Duchovic\cite{Reference480}\\
$^j$4.7$\times$10$^{-10}$ & CVT + RRKM &	\hspace{2mm} & Pilling\cite{Reference466}\\
$^j$4.3--4.7$\times$10$^{-10}$ & $\mu$VT + RRKM, &	\hspace{2mm} & Forst\cite{Reference465}\\
 & CVT + RRKM & \hspace{2mm} & \\
$^j$3.3$\times$10$^{-10}$ & SACM/CT &	\hspace{2mm}  CASPT2/cc-pVDZ$^k$ & Troe and Ushakov\cite{Reference467}\\
$^j$3.2$\times$10$^{-10}$ & VRC-TST &	\hspace{2mm} CASPT2/cc-pVDZ$^k$ & Harding et al.\cite{Reference478}\\
2.0--2.7$\times$10$^{-10}$ & CVT &	\hspace{2mm} MP4/6-31G**, & Hase et al.\cite{Reference481}\\
 & & \hspace{2mm} MRD-CI/6-31G** & \\
2.1$\times$10$^{-10}$ & CVT &	\hspace{2mm} MCSCF-CI/DZP$^l$ & Takahashi et al.\cite{Reference479}\\
\ce{2CH3 -> C2H6} & & & \\[+1mm]
1.8$\times$10$^{-10}$--4.6$\times$10$^{-9}$ & CVT &	\hspace{2mm} CASPT2/ANO-L$^m$ & Li et al.\cite{Reference495}\\
$^j$6.9--8.4$\times$10$^{-11}$ & LTS &	\hspace{2mm} MC & Wardlaw and Marcus\cite{Reference502}\\
8.3$\times$10$^{-11}$ & CVT &	\hspace{2mm} MRD-CI/DZ & Darvesh et al.\cite{Reference496}\\
$^j$7.2$\times$10$^{-11}$ & FTST &	\hspace{2mm} MC & Pesa et al.\cite{Reference501}\\
$^j$5.8--6.7$\times$10$^{-11}$ & $\mu$VT + RRKM &	\hspace{2mm} & Forst\cite{Reference465}\\
$^j$6.3$\times$10$^{-11}$ & CVT &	\hspace{2mm} CAS+1+2/cc-pVDZ & Klippenstein and Harding\cite{Reference498}\\
 & CVT + RRKM & \hspace{2mm} & \\
5.8$\times$10$^{-11}$ & VRC-TST &	\hspace{2mm} CASPT2/cc-pVDZ$^k$ & Klippenstein et al.\cite{Reference500}\\
$^j$5.8$\times$10$^{-11}$ & RRKM + FTST &	\hspace{2mm} & Wagner and Wardlaw\cite{Reference488}\\
$^j$5.6$\times$10$^{-11}$ & CVT &	\hspace{2mm} MRCI+Q/aug-cc-pVTZ & Wang et al.\cite{Reference497}\\
2.0$\times$10$^{-11}$ & CVT &	\hspace{2mm} B3LYP/6-31G** & Lorant et al.\cite{Reference499}\\
\ce{^3CH2 + H2 -> CH3 + H} & & & \\[+1mm]
1.5$\times$10$^{-18}$ & TST & \hspace{2mm} G2M(RCC2)$^n$ & Lu et al.\cite{Reference546}\\
\ce{^3CH2 + ^4N -> H2CN* -> HCN + H} & & & \\[+1mm]
7.9$\times$10$^{-11}$ & quantum dynamics & \hspace{2mm}   MP4SDTQ/6-311++G(3df,3pd)$^o$ &  Herbst et al.\cite{Reference574}\\
\ce{^3CH2 + ^3CH2 -> C2H2 + 2H} & & & \\[+1mm]
1.5$\times$10$^{-10}$ & VRC-TST & CASPT2/aug-cc-pVTZ$^p$ \hspace{2mm}   & Jasper et al.\cite{Reference577}\\
\ce{^3CH2 + CH3 -> C2H4 + H} & & & \\[+1mm]
2.2$\times$10$^{-10}$ & VRC-TST & CASPT2/aug-cc-pVTZ$^p$ \hspace{2mm}   & Jasper et al.\cite{Reference577}\\
\ce{CH + H2 -> CH3} & & & \\[+1mm]
$^j$7.8$\times$10$^{-11}$ & RRKM & \hspace{2mm}  & McIlroy and Tully\cite{Reference576}\\
\ce{CH + H2 -> CH3* -> ^3CH2 + H} & & & \\[+1mm]
3.3$\times$10$^{-11}$ & QCT & \hspace{2mm} MRCI/aug-cc-pVTZ & Mayneris et al.\cite{Reference547}\\
\ce{CH + N -> ^3HCN* -> CN + H} & & & \\[+1mm]
1.2$\times$10$^{-10}$ & quantum dynamics & \hspace{2mm} MRCI+Q/AVTZ & Daranlot et al.\cite{Reference572}\\
\ce{^3NH + ^4N -> N2 + H} & & & \\[+1mm]
1.9$\times$10$^{-11}$ & QCT &	\hspace{2mm} MRCI/aug-cc-pVQZ & Caridade et al.\cite{Reference524}\\
\ce{^3NH + H -> H2 + ^4N} & & & \\[+1mm]
1.5$\times$10$^{-12}$ & CT &	\hspace{2mm} MRCI/aug-cc-pVQZ & Adam et al.\cite{Reference527}\\
1.3--5.2$\times$10$^{-12}$ & CVT &	\hspace{2mm} MP-SAC4/6-311G** & Xu et al.\cite{Reference585}\\
2.0--5.5$\times$10$^{-13}$ & QCT, CVT &	\hspace{2mm} MCQDPT2/6-311++G**$^q$ & Pascual et al.\cite{Reference528}\\
\ce{^2N + H2 -> NH2* -> ^3NH + H} & & & \\[+1mm]
2.5--3.3$\times$10$^{-12}$ & quantum dynamics, QCT & FOCI/TZP & Takayanagi et al.\cite{Reference550}\\
2.7--2.9$\times$10$^{-12}$ & QCT & FOCI/TZP & Kobayashi et al.\cite{Reference551}\\
1.8--2.7$\times$10$^{-12}$ & QCT, CVT &  & Suzuki et al.\cite{Reference552}\\
8.9$\times$10$^{-13}$ & QCT & MRCI/aug-pVTZ & Pederson et al.\cite{Reference567}\\
\hline
\multicolumn{4}{l}{\footnotesize $^a$ Single point energies are based on optimized geometries calculated at the MP2/cc-pVTZ level.} \\
\multicolumn{4}{l}{\footnotesize $^b$ Energy barrier scaled by a factor of 1.174.} \\
\multicolumn{4}{l}{\footnotesize $^c$ Single point energies are based on optimized geometries calculated at the BHandHLYP/6-311G(d,p) level.} \\
\multicolumn{4}{l}{\footnotesize $^d$ Single point energies are based on optimized geometries calculated at the BHandHLYP/cc-pVDZ level.} \\
\multicolumn{4}{l}{\footnotesize $^e$ Energy barrier scaled by a factor of 0.86.} \\
\multicolumn{4}{l}{\footnotesize $^f$ Single point energies are based on optimized geometries calculated at the MP2/6-31G(d) level.} \\
\multicolumn{4}{l}{\footnotesize $^g$ Single point energies are based on optimized geometries calculated at the HF/6-31G(d) level.} \\
\multicolumn{4}{l}{\footnotesize $^h$ Single point energies are based on optimized geometries calculated at the UMP2/6-31G** level.} \\
\multicolumn{4}{l}{\footnotesize $^i$ Single point energies are based on optimized geometries calculated at the B3LYP/cc-pVTZ level.} \\
\multicolumn{4}{l}{\footnotesize $^j$ Values calculated in the high-pressure limit (p $\rightarrow \infty$).} \\
\multicolumn{4}{l}{\footnotesize $^k$ Single point energies are based on optimized geometries calculated at the B3LYP/6-31G* level,} \\
\multicolumn{4}{l}{\footnotesize and corrections are applied at the CAS+1+2+QC/aug-cc-pVTZ level.} \\
\multicolumn{4}{l}{\footnotesize $^l$ CI calculations are based on optimized geometries calculated at the UHF/DZP level.} \\
\multicolumn{4}{l}{\footnotesize $^m$ Single point energies are based on optimized geometries calculated at the CASSCF/ANO-L level.} \\
\multicolumn{4}{l}{\footnotesize $^n$ Single point energies are based on optimized geometries calculated at the B3LYP/6-311++G(3df, 2p) level.} \\
\multicolumn{4}{l}{\footnotesize $^o$ Single point energies are based on optimized geometries calculated at the MP2/6-31G(d,p) level.} \\
\multicolumn{4}{l}{\footnotesize $^p$ Single point energies are based on optimized geometries calculated at the B3LYP/6-311++G(d,p) level.} \\
\multicolumn{4}{l}{\footnotesize $^q$ Single point energies are based on optimized geometries calculated at the FORS-MCSCF(7,6)/6-311++G** level.} \\
\multicolumn{4}{l}{\footnotesize TST: Transition state theory.} \\
\multicolumn{4}{l}{\footnotesize CVT: Canonical variational transition state theory.} \\
\multicolumn{4}{l}{\footnotesize SCT: Small curvature tunneling approximation} \\
\multicolumn{4}{l}{\footnotesize $\mu$OMT: Microcanonical optimized multidimensional tunneling.} \\
\multicolumn{4}{l}{\footnotesize BEBO: bond-energy-bond-order method.} \\
\multicolumn{4}{l}{\footnotesize WTC: Wigner tunneling correction.} \\
\multicolumn{4}{l}{\footnotesize RRKM: Rice-Ramsperger-Kassel-Marcus theory.} \\
\multicolumn{4}{l}{\footnotesize $\mu$VT: Microcanonical variational transition state theory.} \\
\multicolumn{4}{l}{\footnotesize SACM/CT: Statistical adiabatic channel model/classical trajectories approach.} \\
\multicolumn{4}{l}{\footnotesize VRC-TST: Variable reaction coordinate transition state theory.} \\
\multicolumn{4}{l}{\footnotesize LTS: Loose transition state model.} \\
\multicolumn{4}{l}{\footnotesize MC: Monte Carlo simulations.} \\
\multicolumn{4}{l}{\footnotesize FTST: Flexible transition state theory.} \\
\multicolumn{4}{l}{\footnotesize QCT: Quasi-classical trajectory method} \\
\multicolumn{4}{l}{\footnotesize CT: Classical trajectory method} \\
\end{longtable*}

\subsection*{Example Calculation and Computational Methods Comparison}

The \ce{CH4 + H -> CH3 + H2} abstraction reaction has been thoroughly studied both experimentally and theoretically \cite{Reference450,Reference451,Reference452,Reference453,Reference454,Reference455,Reference456,Reference457,Reference458,Reference459,Reference460,Reference461,Reference462,Reference590}. 
A pair of experiments at 298 K place its rate coefficient between 1.7--3.5$\times$10$^{-17}$ cm$^3$s$^{-1}$ \cite{Reference452,Reference450}. However, based on a range of experiments and evaluations over a wider temperature range (300--2000 K), it has been suggested the rate coefficient is closer to 8.2$\times$10$^{-19}$ cm$^3$s$^{-1}$ \cite{Reference451}. Previous theoretical studies have calculated its rate coefficient to be between 1.8$\times$10$^{-21}$ and 2.2$\times$10$^{-18}$ cm$^3$s$^{-1}$ (see Table~\ref{TableTheor}).

The geometry of this reaction progresses as follows: A single H atom approaches a CH$_4$ molecule directly in line with one of its H atoms and its central C atom. The H-C bond in methane then stretches until its H atom bonds with the adjacent H atom. The two newly formed molecules, H$_2$ and CH$_3$ then separate. The geometry of the transition state is depicted in Figure~\ref{Figure3}.

\begin{figure}[!hbtp]
\centering
\includegraphics[width=\linewidth]{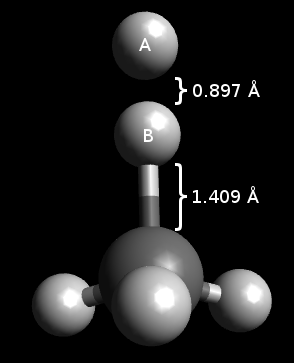}
\caption{Geometry of the conventional transition state for \ce{CH4 + H -> CH3 + H2} at the BHandHLYP/aug-cc-pVDZ level of theory. In the reactant state, hydrogen B is 1.09$\AA$ from the central carbon. In the product state hydrogen B is 0.754 $\AA$ from hydrogen A.}
\label{Figure3}
\end{figure}

In Figure~\ref{Figure3}, we show $\Delta G_{GT}(298.15 K,s)$ with the reaction coordinate representing the C-H bond distance. At the BHandHLYP/aug-cc-pVDZ level of theory, the maximum $\Delta$G occurs at a C-H distance of 1.44$\AA$, which is slightly farther along the reaction coordinate than the conventional transition state (1.409$\AA$).

\begin{figure}[!hbtp]
\centering
\includegraphics[width=\linewidth]{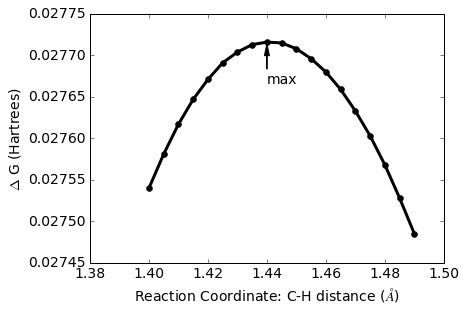}
\caption{Gibbs free energy difference as a function of the reaction coordinate (C-H bond distance) for \ce{CH4 + H -> CH3 + H2} at the BHandHLYP/aug-cc-pVDZ level of theory. The maximum $\Delta$G occurs at 1.44$\AA$. $\Delta$G is calculated as the Gibbs free energy at the reaction coordinate minus the Gibbs free energy of the reactants placed 100$\AA$ apart.}
\label{Figure3}
\end{figure}

We calculate the rate coefficient for this reaction 
using 6 different computational methods, and display them in Table~\ref{TableValidate}.

\begin{table}[t]
\centering
\caption{Calculated rate coefficients for \ce{CH4 + H -> CH3 + H2} using 6 different computational methods with the aug-cc-pVDZ basis set. Experimental and suggested values range from 8.2$\times$10$^{-19}$ to 3.5$\times$10$^{-17}$ cm$^3$s$^{-1}$ \cite{Reference452,Reference450,Reference451}. The error factor is the multiplicative or divisional factor from the nearest experimental or suggested value; the error factor is 1 if the calculated value is within the range of experimental or suggested values. Rate coefficients have units cm$^{3}$s$^{-1}$. \label{TableValidate}}
\begin{tabular}{ccc}
\\
\multicolumn{1}{c}{Computational Method} &
\multicolumn{1}{c}{k(298)} &
\multicolumn{1}{c}{Error factor}\\ \hline \\[-2mm]
HF & 3.5$\times$10$^{-25}$ & 2$\times$10$^{6}$\\[+1mm]
M06-2x & 5.7$\times$10$^{-20}$ & 14\\
CCSD & 9.9$\times$10$^{-20}$ & 8\\
BHandHLYP & 8.1$\times$10$^{-18}$ & 1\\
CAM-B3LYP & 3.1$\times$10$^{-17}$ & 1\\
B3LYP & 3.9$\times$10$^{-16}$ & 11\\
\hline
\end{tabular}
\end{table}

Rate coefficient calculations using BHandHLYP and CAM-B3LYP methods are within the experimental and suggested range. Calculations using HF grossly over-estimate the energy barrier, and provide a rate coefficient several orders of magnitude lower than the experimental and suggested range. M06-2x and CCSD methods also provide values lower than the experimental and suggested range, however only by approximately an order of magnitude. Calculations using B3LYP under-estimate the energy barrier, leading to a value approximately an order of magnitude higher than the experimental and suggested range.

The calculated rate coefficient using the BHandHLYP functional sits in the middle of the range of experimental and suggested values, whereas he rate coefficient calculated using CAM-B3LYP sits within $\sim$10\% of the experimental value at the high end of the range. The BHandLYP result may provide the best compromise between the experimental and suggested rate coefficients for this reaction.

Finally, CCSD provides a fairly accurate rate coefficient within a factor of 8 of the suggested value.
 
 %endogenous source of nucleobases and amino acids on the prebiotic Earth and on Saturn's moon Titan, is from atmospheric HCN dissolving into (sometimes transient) liquid water on the surface for aqueous reaction \cite{2017JGRE..122..432H,Reference441}.

%Simple lifeforms comprised of catalytic RNA molecules likely preceded the emergence of DNA- and protein-based life \cite{1986Natur.319..618G,Reference148}. The first RNA polymers chained together out of building blocks comprised of a characteristic nucleobase, D-ribose, and phosphate. Endogenous production of nucleobases on the early Earth would 

%The abiotic production of nucleobases requires a reactive source of nitrogen, typically HCN or NH$_3$ \cite{Reference46,Reference438}. 

\subsection*{Theoretical Case Studies}

\subsubsection*{Case Study 1: \ce{H2CN + H -> HCN + H2}}

Tomeczek and Grado{\'n}\cite{Reference446} used published chemical compositions of the flames of CH$_4$ and O$_2$ + N$_2$ at 2500 to 1850 K to suggest a temperature-independent rate coefficient for \ce{H2CN + H -> HCN + H2}. They suggest the value 8.3$\times$10$^{-11}$ cm$^3$s$^{-1}$ for this reaction. However, they note that this does not include the effects of an energy barrier. Another way to state this is, they suggest a value for the entropic component of this reaction, but not the energetic component.

We find no previous theoretical reaction rate coefficients for \ce{H2CN + H -> HCN + H2}.

This reaction occurs on the singlet and triplet PES's. There is no energy barrier for this reaction on the singlet PES. Conversely on the triplet PES, where excited $^3$HCN is produced, the effects of the energy barrier are significant. 

On the singlet surface, we calculate the reaction rate coefficient at the BHandHLYP/aug-cc-pVDZ level of theory to be 1.8$\times$10$^{-11}$ cm$^3$s$^{-1}$. This is less than a factor of 5 larger than the experimental value for the barrierless reaction. 

On the triplet surface, the reaction rate coefficient is too small to consider in this study ($k <$ 10$^{-21}$ cm$^3$s$^{-1}$).

\subsubsection*{Case Study 2: \ce{2H2CN -> HCN + H2CNH}}

Horne and Norrish\cite{Reference447} calculated the experimental reaction rate coefficient for \ce{2H2CN -> HCN + H2CNH} at 300 K by monitoring the decay of H$_2$CN. The assumption they made was that \ce{2H2CN -> HCN + CH2NH} is the dominant decay pathway of H$_2$CN. The value they obtained was in the range of 3.3--8.3$\times$10$^{-12}$ cm$^3$s$^{-1}$.

No theoretical reaction rate coefficients for \ce{2H2CN -> HCN + H2CNH} have been previously published.

We find a direct reaction pathway on the singlet PES that has no energy barrier. However, the simulations did not converge beyond a N-H bond distance of 1.95 $\AA$ and the Gibbs maximum was not found. However, choosing the reaction coordinate at a N-H bond distance of 1.95 $\AA$ for the calculation provides us with a lower bound estimate of the rate coefficient, which we calculate to be 3.7$\times$10$^{-14}$ cm$^3$s$^{-1}$. This value is a factor of 89 smaller than the closest experimental value. The discrepancy between the theoretical and experimental values is expected to be due to these convergence issues.

A higher energy reaction pathway involving two ground state H$_2$CN molecules exists on the triplet surface, however, this reaction produces excited $^3$HCN and is likely much less efficient than the singlet case.

%In the multi-step reaction \ce{CH3 + ^2N -> ^1H3CN* -> ^1H2CNH* -> H2CN + H}, we include the decay of H$_2$CNH into H$_2$CN + H. For consistency, we include the decay of H$_2$CNH into H$_2$CN + H in this reaction as well, i.e. \ce{2H2CN -> HCN + H2CNH* -> HCN + H2CN + H}.

%Alternate decay pathways of H$_2$CN could also  include \ce{H2CN -> HCN + H}, \ce{H2CN + H -> HCN + H2}, \ce{H2CN + H -> ^3H2CNH}, \ce{H2CN + H -> ^1H2CNH} and \ce{H2CN + H -> H3CN}. Out of the above three reactions, the dominant decay pathways should correspond to the ones with the highest rate coefficients. We calculate the reaction rate coefficients for the above five pathways at the BHandHLYP/aug-cc-pVDZ level of theory to be 1.7$\times$10$^{-11}$ s$^{-1}$, 1.9$\times$10$^{-11}$ cm$^3$ s$^{-1}$, 1.0$\times$10$^{-16}$ cm$^3$s$^{-1}$, 8.9$\times$10$^{-10}$ cm$^3$s$^{-1}$, and 3.5$\times$10$^{-13}$ cm$^3$ s$^{-1}$, respectively. Thus, H$_2$CN is most frequently going to decay into HCN and CH$_2$NH through the pathways \ce{H2CN + H -> HCN + H}, \ce{H2CN -> HCN + H}, and \ce{H2CN + H -> H2CNH}.

\subsubsection*{Case Study 3: \ce{CH4 + N -> products}}

Several experiments have measured the rate coefficient of \ce{CH4 + ^2N -> products} by monitoring the decay of $^2$N in the presence of CH$_4$ at 295--300 K \cite{Reference579,Reference581,Reference555,Reference559}. The measured values range from 3.0--5.4$\times$10$^{-12}$ cm$^3$s$^{-1}$. Herron\cite{Reference549} reviewed these experiments and recommended a value of k(298) = 4.0$\times$10$^{-12}$ cm$^3$s$^{-1}$. Umemoto et al.\cite{Reference580} measured the product yields of H and $^3$NH in similar experiments to suggest branching ratios for \ce{CH4 + ^2N -> H2CNH + H} and \ce{CH4 + ^2N -> CH3 + ^3NH} to be 0.8$\pm$0.2 and 0.3$\pm$0.1, respectively.
Multiplying with these branching ratios, the experimental rate coefficients for \ce{CH4 + ^2N -> H2CNH + H} range from 2.4--4.5$\times$10$^{-12}$ cm$^3$s$^{-1}$ and the experimental rate coefficients for \ce{CH4 + ^2N -> CH3 + ^3NH} range from 0.9--1.7$\times$10$^{-12}$ cm$^3$s$^{-1}$.

Takayanagi et al.\cite{Reference579} used CVT at the CASSCF(5,5)/6-311G** level of theory to calculate the rate coefficient of \ce{CH4 + N -> products} to be 8.5$\times$10$^{-14}$ cm$^3$s$^{-1}$. They note that their disagreement between experimental and theoretical values is due to the CASSCF calculations estimating too large a barrier. Ouk et al.\cite{Reference582} used TST + WTC at the MRCI+P+Q/aug-cc-pVTZ level of theory to calculate the rate coefficient, and obtained a value closer to the experimental values at 6.8$\times$10$^{-12}$ cm$^3$s$^{-1}$. They confirm the results from experiment that suggests a small barrier exists, although no barrier is found using the CCSD(T) and B3LYP levels of theory \cite{Reference583}. The experimental barrier has a height of 6.3 kJ mol$^{-1}$ \cite{Reference579}.

\begin{figure}[!hbtp]
\centering
\includegraphics[width=\linewidth]{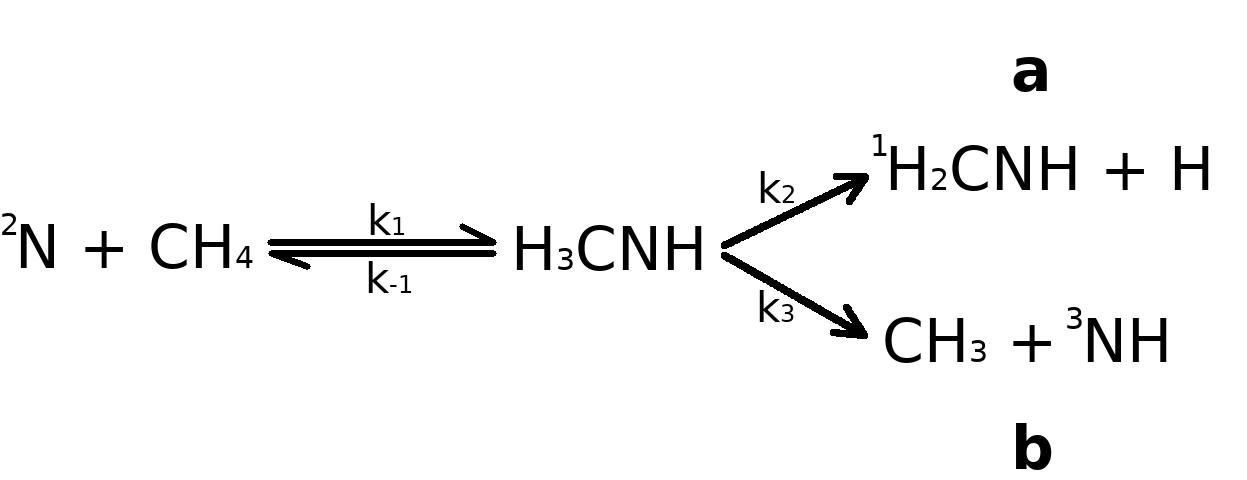}
\caption{Mechanistic models for the production of ({\bf a}) $^1$H$_2$CNH + H and ({\bf b}) CH$_3$ + $^3$NH from $^2$N + CH$_4$ on the doublet potential energy surface. \label{FigureX}}
\end{figure}

In this case study, we analyze the two main branches for the reaction \ce{CH4 + ^2N -> products}. The mechanistic model for these reactions is shown in Figure~\ref{FigureX}.

Similar to Balucani et al.\cite{Reference583}, we find no barrier for the \ce{CH4 + ^2N -> H3CNH} reaction step. We also run into convergence problems when stretching the C-N bond distance farther than 2.82 $\AA$. Therefore, we are unable to find a Gibbs maximum, and instead choose the reaction coordinate at 2.82 $\AA$ for our CVT calculation of $k$. Past theoretical works found the transition state to be at a C-N bond distance of 2.26--2.45 $\AA$ \cite{Reference583,Reference584,Reference582}.

We find the barrierless reaction rate coefficient of \ce{CH4 + ^2N -> H3CNH} at the BHandHLYP/aug-cc-pVDZ level of theory to be 6.0$\times$10$^{-10}$ cm$^3$s$^{-1}$, a value two orders of magnitude greater than the closest experimental value. Given that there is likely a small barrier of $\sim$6.3 kJ mol$^{-1}$ for this reaction, we insert this experimental barrier into the calculation for $k$, and obtain a value of 4.7$\times$10$^{-11}$ cm$^3$s$^{-1}$. This is only a factor of 9 larger than the nearest experimental value.

There are several decay pathways for the H$_3$CNH molecule (e.g. \cite{Reference583}). Nevertheless, we calculate the upper bound for the rate coefficients of \ce{CH4 + ^2N -> H2CNH + H} and \ce{CH4 + ^2N -> CH3 + ^3NH} by assuming H$_3$CNH only decays through these two dominant pathways.

The steady-state solution of the kinetic rate equations for the above mechanistic models lead to the overall rate constants for a) \ce{CH4 + ^2N -> ^1H2CNH + H} and b) \ce{CH4 + ^2N -> CH3 + ^3NH} on the doublet PES.

\begin{equation}
k_a = \frac{k_{1} k_{2}}{k_{-1} + k_{2} + k_3}
\end{equation}

\begin{equation}
k_b = \frac{k_{1} k_{3}}{k_{-1} + k_{2} + k_3}
\end{equation}

The values of the reaction rate constants at the BHandHLYP/aug-cc-pVDZ level of theory are listed in Table~\ref{TableX}.

\begin{table}[t]
\centering
\caption{Calculated overall rate coefficients for \ce{CH4 + ^2N -> ^1H2CNH + H} and \ce{CH4 + ^2N -> CH3 + ^3NH}, as well as the intermediate forward and reverse rate coefficients which were used in the calculations. In all simulations, the BHandHLYP method was used with the aug-cc-pVDZ basis set. When using a branching ratio of 0.8 for \ce{CH4 + ^2N -> ^1H2CNH + H} \cite{Reference580}, experiments from 295--300 K provide k$_a$ values between 2.4 and 4.5$\times$10$^{-12}$ cm$^3$s$^{-1}$. Similarly when using a branching ratio of 0.3 for \ce{CH4 + ^2N -> CH3 + ^3NH}, the same experiments provide k$_b$ values between 0.9--1.7$\times$10$^{-12}$ cm$^3$s$^{-1}$. First-order rate coefficients have units s$^{-1}$. Second-order rate coefficients have units cm$^{3}$s$^{-1}$. \label{TableX}}
\begin{tabular}{cc}
\\
\multicolumn{1}{c}{Rate coefficient} &
\multicolumn{1}{c}{k(298)}\\ \hline \\[-2mm]
k$_a$ & 4.7$\times$10$^{-11}$ \\[+1mm]
k$_b$ & 5.8$\times$10$^{-29}$ \\
k$_{1}$ & $^a$4.7$\times$10$^{-11}$ \\
k$_{-1}$ & 3.0$\times$10$^{-66}$ \\
k$_{2}$ & 4.9$\times$10$^{-13}$ \\
k$_{3}$ & 6.0$\times$10$^{-31}$ \\
\hline
\multicolumn{2}{l}{\footnotesize $^a$ Simulations did not converge beyond a C-N bond distance of} \\
\multicolumn{2}{l}{\footnotesize 2.82$\AA$; therefore the rate coefficient is a lower bound.} \\
\end{tabular}
\end{table}

We calculate the overall rate constant for \ce{CH4 + ^2N -> H3CNH* -> ^1H2CNH + H} to be the same as the rate constant for \ce{CH4 + ^2N -> H3CNH}. This means the first step is the rate-limiting step. We calculate the value for k$_a$ to be 4.7$\times$10$^{-11}$ cm$^{3}$s$^{-1}$, which is approximately a factor of 10 smaller than the experimental values. We find the overall rate coefficient for \ce{CH4 + ^2N -> H3CNH* -> CH3 + ^3NH} to be inefficient (k$_b$ = 5.8$\times$10$^{-29}$ cm$^{3}$s$^{-1}$). This is several orders of magnitude smaller than the rate coefficients suggested by the experimental branching ratios \cite{Reference580}. However, some theoretical branching ratios for \ce{CH4 + ^2N -> H3CNH* -> CH3 + ^3NH} are as low as 0.01 \cite{Reference582}, and in every case, \ce{CH4 + ^2N -> H3CNH* -> ^1H2CNH + H} is the dominant product. Considering all this, we do not include the inefficient \ce{CH4 + ^2N -> H3CNH* -> CH3 + ^3NH} reaction in our network.

\subsubsection*{Case Study 4: \ce{CH3 + N -> products}}

Stief et al.\cite{Reference444} experimentally calculated the overall reaction rate constant for \ce{CH3 + N -> products} at 298 K to be 8.6$\times$10$^{-11}$ cm$^3$s$^{-1}$ by monitoring the decay of reactants CH$_3$ and N in a volume. Marston et al.\cite{Reference442} suggest the three possible branches for the reaction of \ce{CH3 + N -> products} are:

\begin{align*}
\ce{CH3 + N -> H2CN + H},
\end{align*}
\begin{align*}
\ce{CH3 + N -> HCN + H2},
\end{align*}
and
\begin{align*}
\ce{CH3 + N -> HCN + 2H}.
\end{align*}

Marston et al.\cite{Reference442} monitored the production of H$_2$, and H in experiments reacting CH$_3$ and N, and calculated the above reaction branching ratios to be approximately 0.9, 0.1, and 0 respectively. This suggests a preference for the \ce{CH3 + N -> H2CN + H} pathway by approximately an order of magnitude over the \ce{CH3 + N -> HCN + H2} pathway. It must be noted that in performing this calculation, Marston et al.\cite{Reference442} assumed that H$_2$ and H were solely generated through the above pathways. They caution the reader that it is also possible that these products formed through the H$_2$CN intermediate.

Miller and Bowman\cite{Reference445} suggested the rate coefficient of \ce{CH3 + N -> H2CN + H} to be 5.0$\times$10$^{-11}$ cm$^3$s$^{-1}$ based on thermodynamic calculations.

There are two main PES's that the CH$_3$ + N reaction evolves on: the triplet and the singlet surfaces. The quintet surface is also possible, however this reaction is much higher in energy and therefore much less likely to occur \cite{Reference464}. 
Both the ground state nitrogen atom (i.e. $^4$N) and the excited nitrogen atom (i.e. $^2$N) can react with CH$_3$ on the triplet PES. Only the excited state nitrogen atom can react with CH$_3$ on the singlet PES.

%Both $^4$N and $^2$N are formed in the UV photodissociation of N$_2$ in even amounts \cite{2017ApJ...850...48S}. $^2$N also survives for several hours before radiatively decaying into the ground state as this is a spin-forbidden process. The collisional deexcitation of $^2$N by N$_2$ is also a slow process, thus the main atmospheric loss of $^2$N is from chemical reactions \cite{2013ApJS..204...20D}.

%The ground states of HCN, and H$_2$CN + H are in the singlet and triplet spin states, respectively. Therefore the spin-allowed reaction channels for these reactions are \ce{CH3 + ^4N -> ^3H3CN* -> H2CN + H} and \ce{CH3 + ^2N -> ^1H3CN* -> HCN + H2}, respectively. Producing HCN on the triplet PES and H$_2$N + H on the singlet PES may be possible, however doing so requires forming HCN and H$_2$CN in their higher energy, excited states, and is less probable. Spin-forbidden processes are also possible, i.e., evolving CH$_3$ + $^4$N into HCN + H$_2$ by passing through the conical intersection between the triplet and singlet PES's. However, such processes are inherently less probable than spin-allowed processes.

A computational study of the \ce{CH3 + N -> products} reaction shows a preference for the \ce{CH3 + ^4N -> H2CN + H} pathway \cite{Reference464}. This study finds the \ce{CH3 + ^2N -> HCN + H2} channel to be negligible. Cimas and Largo\cite{Reference464} suggest that the HCN measured in experiments by Marston et al.\cite{Reference442} formed through the H$_2$CN intermediate, via reaction equations~\ref{Eq8}--\ref{Eq10}. Chiba and Yoshida\cite{Reference539} alternatively suggest that HCN + H$_2$ may form through the triplet-singlet spin-forbidden process.

Alves et al.\cite{Reference463} and Cimas and Largo\cite{Reference464} analyzed \ce{CH3 + N -> products} theoretically using quantum computational simulations at the CCSD(T)/CBS and CCSD(T)/cc-pVTZ levels of theory, and calculated its reaction rate coefficients to be 1.93$\times$10$^{-10}$ cm$^3$s$^{-1}$ and 9.1$\times$10$^{-12}$ cm$^3$s$^{-1}$, respectively.

In this case study, we analyze the three suggested main branches for \ce{CH3 + N -> products} using CVT (see the methods section for full details). Computational studies show that \ce{CH3 + ^4N -> products} reactions first proceed through a barrierless reaction to H$_3$CN on the triplet surface \cite{Reference464,Reference463}. We confirm this barrierless reaction (\ce{CH3 + ^4N -> ^3H3CN}) and calculate its rate coefficients at the BHandHLYP/aug-cc-pVDZ level of theory to be 3.3$\times$10$^{-11}$ cm$^3$s$^{-1}$. This result is less than a factor of 3 smaller than the experimental result (8.6$\times$10$^{-11}$ cm$^3$s$^{-1}$ \cite{Reference444}). Our calculated rate coefficient is also within a factor of 3 of the calculated value by Cimas and Largo\cite{Reference464} at the CCSD(T)/cc-pVTZ level of theory (9.1$\times$10$^{-12}$ cm$^3$s$^{-1}$).

%We also compute the barrierless reaction on the singlet PES \ce{CH3 + ^2N -> ^1H3CN}. The rate coefficient for this reaction at the BHandHLYP/aug-cc-pVDZ level of theory is 1.0$\times$10$^{-10}$ cm$^3$s$^{-1}$.

We do not find a direct reaction pathway on the singlet or triplet surface to \ce{CH3 + N -> HCN + 2H}.

\begin{figure}[!hbtp]
\centering
\includegraphics[width=\linewidth]{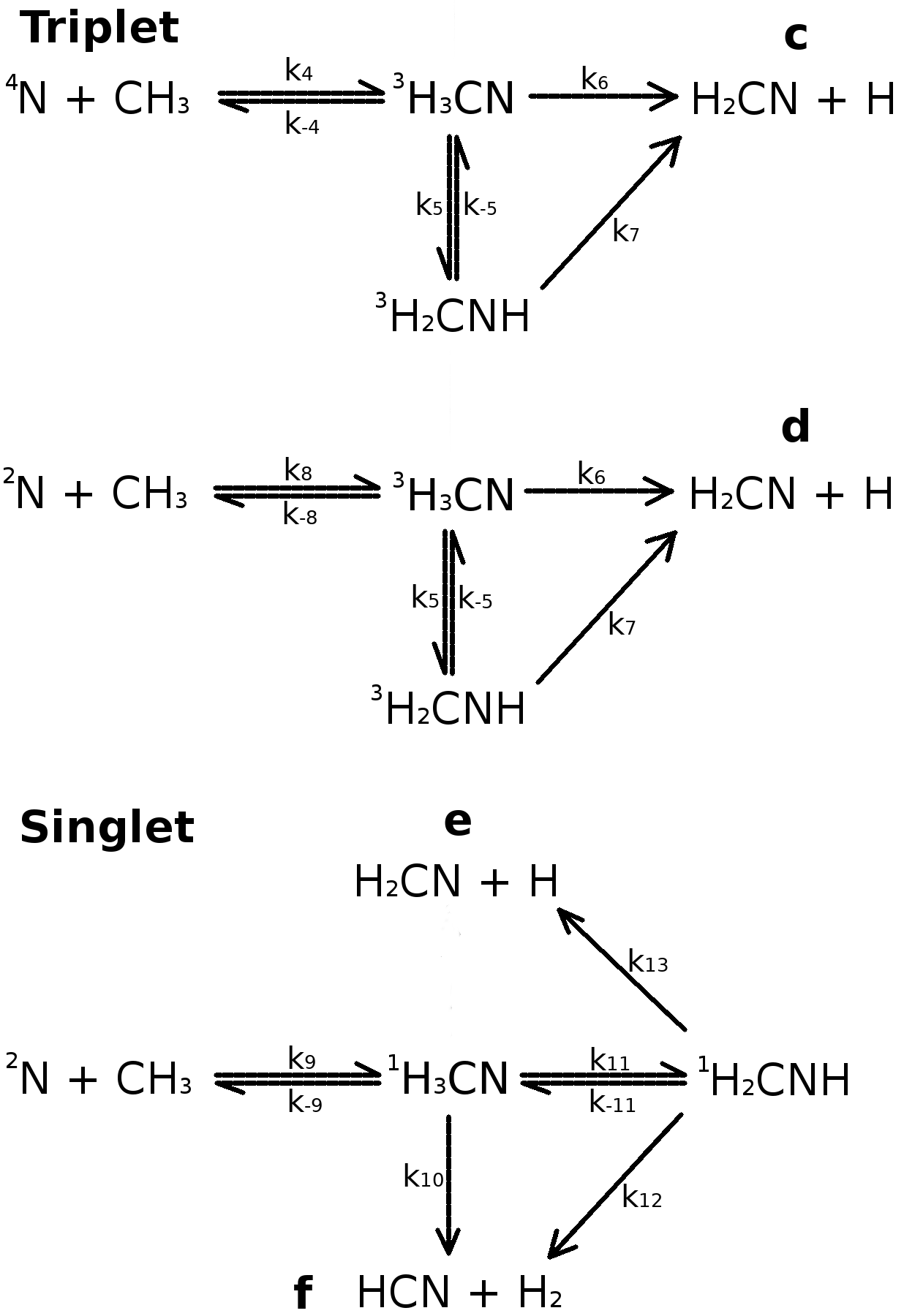}
\caption{Mechanistic models for the production of H$_2$CN + H on the triplet surface from reactants ({\bf c}) $^4$N + CH$_3$ and ({\bf d}) $^2$N + CH$_3$, and production of ({\bf e}) H$_2$CN + H and ({\bf f}) HCN + H$_2$ on the singlet surface, from CH$_3$ + $^2$N. \label{NCH3-model}}
\label{Figure4}
\end{figure}

We display the mechanistic models for forming H$_2$CN + H, and HCN + H$_2$ from CH$_3$ + N in Figure~\ref{Figure4}. These mechanistic models are similar to that used in Alves et al.\cite{Reference463}. Note we do not analyze spin-forbidden processes in these models.

%Experiments give an overall reaction rate constant for \ce{CH3 + N -> products} of 8.6$\times$10$^{-11}$ cm$^3$s${-1}$ \cite{Reference444}, and branching ratios of 0.85, 0.10, and 0.05, for $H_2$CN + H, HCN + H$_2$ and HCN + 2H, respectively \cite{Reference442}. Thus, if these three sets of products did not form through any intermediate other than NCH$_3$, the experimental reaction rate coefficients would be 7.3$\times$10$^{-11}$ cm$^3$s${-1}$, 8.6$\times$10$^{-12}$ cm$^3$s${-1}$, and 4.3$\times$10$^{-12}$ cm$^3$s${-1}$, respectively.

%Furthermore, as will be seen in the following calculation, the \ce{CH3 + N -> HCN + H2} spin-forbidden process has a very slow reaction rate coefficient, and is unlikely to be the source of HCN from the CH$_3$ + N reactants.

H$_2$CN + H can form directly from $^3$H$_3$CN, or after isomerization from the intermediate $^3$H$_2$CNH. Similarly, HCN + H$_2$ can form directly from $^1$H$_3$CN, or from the intermediate $^1$H$_2$CNH. On the singlet surface, we find H$_2$CN + H forms from the intermediate $^1$H$_2$CNH, however we do not find a pathway from $^1$H$_3$CN. We find a smooth decrease in Gibbs free energy along the MEP for the reaction \ce{^1H3CN -> ^1H2CNH}, suggesting this reaction has neither an energy barrier nor an entropic barrier. We estimate the rate coefficient for this reaction by choosing the reactant geometry as the transition state. We find the overall rate coefficients for \ce{CH3 + N -> H2CN + H} and \ce{CH3 + N -> H2CN + H} to be insensitive to this intermediate rate coefficient by varying the latter's value by over 10 orders of magnitude in both directions. The optimization of $^1$H$_3$CN does not converge, therefore we use a reactant geometry close to $^1$H$_3$CN that has vibrational modes for HCN + H$_2$ and $^1$H$_2$CNH. In any case, we find the values of k$_c$ and k$_d$ are independent of the $^1$H$_3$CN geometry.

The steady-state solution of the kinetic rate equations for the above mechanistic models give us the overall rate constants for c) \ce{CH3 + ^4N -> H2CN + H} and d) \ce{CH3 + ^2N -> H2CN + H} on the triplet surface, and e) \ce{CH3 + ^2N -> H2CN + H} and f) \ce{CH3 + ^2N -> HCN + H2} on the singlet surface.

\begin{equation}
k_c = \frac{k_4}{A}\left(k_6 + \frac{k_5 k_7}{k_{-5} + k_7}\right).
\end{equation}

\begin{equation}
k_d = \frac{k_8}{B}\left(k_6 + \frac{k_5 k_7}{k_{-5} + k_7}\right).
\end{equation}

\begin{equation}
k_e = \frac{k_9}{C}\left(\frac{k_{11} k_{13}}{k_{-11} + k_{12} + k_{13}}\right).
\end{equation}

\begin{equation}
k_f = \frac{k_9}{C}\left(k_{10} + \frac{k_{11} k_{12}}{k_{-11} + k_{12} + k_{13}}\right).
\end{equation}

\begin{equation}
A = k_{-4} + k_5 + k_6 - \frac{k_{-5} k_5}{k_{-5} + k_7}.
\end{equation}

\begin{equation}
B = k_{-8} + k_5 + k_6 - \frac{k_{-5} k_5}{k_{-5} + k_7}.
\end{equation}

\begin{equation}
C = k_{-9} + k_{10} + k_{11} - \frac{k_{-11} k_{11}}{k_{-11} + k_{12} + k_{13}}.
\end{equation}

The values of the reaction rate constants using the BHandHLYP method and aug-cc-pVDZ basis set are listed in Table~\ref{Table4}.

\begin{table}[t]
\centering
\caption{Calculated overall rate coefficients for \ce{CH3 + ^4N -> H2CN + H} and \ce{CH3 + ^2N -> H2CN + H} on the triplet surface, and \ce{CH3 + ^2N -> H2CN + H} and \ce{CH3 + ^2N -> HCN + H2} on the singlet surface, as well as the intermediate forward and reverse rate coefficients which were used in the calculations. In all simulations, the BHandHLYP method was used with the aug-cc-pVDZ basis set. To reduce computational time, forward and reverse rate coefficients for k$_5$, k$_7$, and k$_{12}$ were calculated with the transition state at the classical location (the saddle point) instead of the variational location. We find the overall rate coefficients k$_c$, k$_d$ and k$_e$ to be insensitive to changes in these intermediate coefficients of over 10 orders of magnitude. k$_f$ is also insensitive to increases in k$_{12}$ by over 10 orders of magnitude, however, decreasing k$_{12}$ directly decreases k$_f$. Because we do not consider reactions with rate coefficients lower than 10$^{-21}$ cm$^3$s$^{-1}$, we make no attempt to increase the accuracy of k$_f$. Experiments at 298 K provide a k$_c$ value of 7.7$\times$10$^{-11}$ and a k$_d$ value of 8.6$\times$10$^{-12}$ cm$^3$s$^{-1}$ \cite{Reference442,Reference444}. First-order rate coefficients have units s$^{-1}$. Second-order rate coefficients have units cm$^{3}$s$^{-1}$. \label{Table4}}
\begin{tabular}{cc}
\\
\multicolumn{1}{c}{Rate coefficient} &
\multicolumn{1}{c}{k(298)}\\ \hline \\[-2mm]
k$_c$ & 3.3$\times$10$^{-11}$\\[+1mm]
k$_d$ & 1.0$\times$10$^{-10}$ \\
k$_e$ & 3.1$\times$10$^{-11}$ \\
k$_f$ & 2.3$\times$10$^{-28}$ \\
k$_4$ & 3.3$\times$10$^{-11}$ \\
k$_{-4}$ & 1.3$\times$10$^{-32}$ \\
k$_5$ & 1.1$\times$10$^{-19}$ \\
k$_{-5}$ & 8.2$\times$10$^{-11}$ \\
k$_6$ & 3.0$\times$10$^{-12}$ \\
k$_7$ & 5.6$\times$10$^{-7}$ \\
k$_8$ & 1.0$\times$10$^{-10}$ \\
k$_{-8}$ & 3.0$\times$10$^{-51}$ \\
k$_9$ & 3.1$\times$10$^{-11}$ \\
k$_{-9}$ & 6.3$\times$10$^{-41}$ \\
k$_{10}$ & 1.5$\times$10$^{-14}$ \\
$^a$k$_{11}$ & 1.9$\times$10$^{13}$ \\
k$_{-11}$ & 3.9$\times$10$^{-57}$ \\
k$_{12}$ & 2.0$\times$10$^{-56}$ \\
k$_{13}$ & 2.9$\times$10$^{-39}$ \\
\hline
\multicolumn{2}{l}{\footnotesize $^a$ No energy or entropic barrier. Transition state chosen at} \\
\multicolumn{2}{l}{\footnotesize reactant geometry ($^1$H$_3$CN).} \\
\end{tabular}
\end{table}

The rate coefficients of k$_c$, k$_d$, and k$_e$ are equivalent to those of k$_4$, k$_8$, and k$_9$, respectively. Thus the rate-limiting steps for these reactions are the first steps, i.e. \ce{CH3 + N -> H3CN}.

On the triplet surface, or theoretical value of k$_c$ at the BHandHLYP/aug-cc-pVDZ level of theory is a factor of 1.5--2.5 smaller than the values calculated in experiments and suggested by thermodynamics \cite{Reference442,Reference444,Reference445}.

On the singlet surface, there is a strong preference to produce H$_2$CN + H over HCN. The rate coefficient of \ce{CH3 + ^2N -> HCN + H2} is less than 10$^{-21}$ cm$^{3}$s$^{-1}$, therefore we do not include this reaction in our network.

%In future work, we will use the rate coefficients calculated in this paper and a non-equilibrium chemical kinetics model to try to account for the HCN produced in experiments by \citet{Reference442}.

\subsubsection*{Case Study 5: \ce{CH2 + H <-> CH3* <-> CH + H2}}

There are three spin configurations for this reaction. \ce{^3CH2 + H -> CH3* -> CH + H2} and \ce{^1CH2 + H -> CH3* -> CH + H2} occur on the doublet PES, and \ce{^3CH2 + H -> ^4CH + H2} occurs on the quartet PES.

Several experiments have calculated the reaction rate coefficient for \ce{^3CH2 + H -> CH + H2} at 285--300 K \cite{Reference503,Reference504,Reference505,Reference506,Reference507,Reference508}. Although methodology differs between experiments, they generally involve monitoring the decay of $^3$CH$_2$. The experimental values are as low as 8.3$\times$10$^{-11}$ cm$^3$s$^{-1}$ and as high as 2.7$\times$10$^{-10}$ cm$^3$s$^{-1}$. Two studies have reviewed a variety of experiments at a range of temperatures and suggested values of 2.7$\times$10$^{-10}$ cm$^3$s$^{-1}$ \cite{Reference509} and 2.0$\times$10$^{-10}$ cm$^3$s$^{-1}$ \cite{Reference451}.

Although no experiments have been performed for the reaction of \ce{^1CH2 + H -> CH + H2}, Tsang and Hampson\cite{Reference509} suggest the value should be near 5.0$\times$10$^{-11}$ cm$^3$s$^{-1}$ based on thermodynamics.

To date there have been no published theoretical reaction rate coefficients for any spin configuration of this reaction.

On the quartet PES, we find the reaction proceeds directly \ce{^3CH2 + H -> ^4CH + H2}. We calculate the rate coefficient for this reaction at the BHandHLYP/aug-cc-pVDZ level of theory to be $\sim$ 10$^{-25}$ cm$^3$s$^{-1}$, which is too inefficient to consider in this network.

\begin{figure}[!hbtp]
\centering
\includegraphics[width=\linewidth]{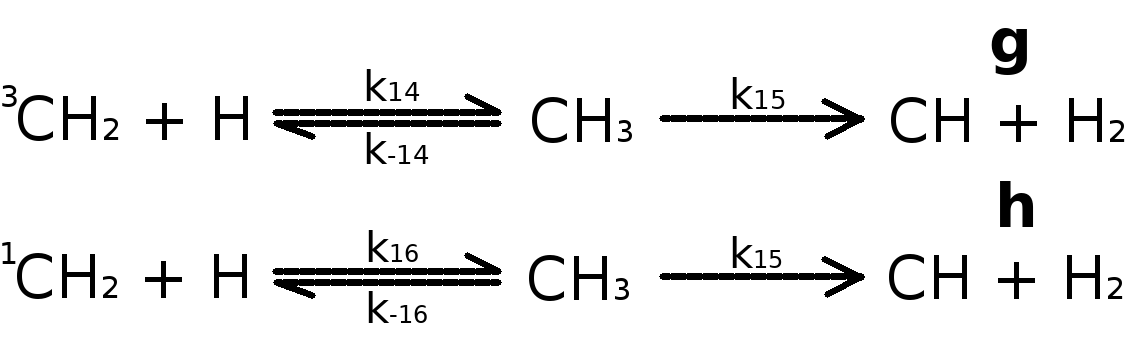}
\caption{Mechanistic models for the production of CH + H$_2$ from $^3$CH$_2$ + H and $^1$CH$_2$ + H.}
\label{Figure5}
\end{figure}

The doublet PES reactions proceed through the CH$_3$ intermediate. The mechanistic model for these reactions is shown in Figure~\ref{Figure5}.

Although there are many reaction pathways for the CH$_3$ molecule, we calculate the upper bound for the rate constants of \ce{^3CH2 + H -> CH3* -> CH + H2} and \ce{^1CH2 + H -> CH3* -> CH + H2} by assuming all CH$_3$ reacts to form CH + H$_2$.

The steady-state solutions of the kinetic rate equations for these mechanistic models give us the overall rate constants for \ce{^3CH2 + H -> CH + H2} and \ce{^1CH2 + H -> CH + H2}.

\begin{equation}
k_g = \frac{k_{15} k_{14}}{k_{-14} + k_{15}}
\end{equation}

\begin{equation}
k_h = \frac{k_{15} k_{16}}{k_{-16} + k_{15}}
\end{equation}

The values of the reaction rate constants using the BHandHLYP method and aug-cc-pVDZ basis set are listed in Table~\ref{Table5}.

\begin{table}[t]
\centering
\caption{Calculated overall rate coefficient for \ce{^3CH2 + H -> CH + H2}, and \ce{^1CH2 + H -> CH + H2}, as well as the intermediate forward and reverse rate coefficients which were used in the calculations. In all simulations, the BHandHLYP method was used with the aug-cc-pVDZ basis set. Experiments at 298 K provide k$_g$ values between 8.3$\times$10$^{-11}$ and 2.7$\times$10$^{-10}$ cm$^3$s$^{-1}$ \cite{Reference503,Reference504,Reference505,Reference506,Reference507,Reference508}. 
A value of 5.0$\times$10$^{-11}$ $^3$s$^{-1}$ is suggested for k$_h$ \cite{Reference509}. First-order rate coefficients have units s$^{-1}$. Second-order rate coefficients have units cm$^{3}$s$^{-1}$. \label{Table5}}
\begin{tabular}{ccc}
\\
\multicolumn{1}{c}{Rate coefficient} &
\multicolumn{1}{c}{k(298)}\\ \hline \\[-2mm]
k$_g$ & 4.3$\times$10$^{-10}$ \\[+1mm]
k$_h$ & 8.4$\times$10$^{-11}$ \\
k$_{14}$ & 5.6$\times$10$^{-10}$ \\
k$_{-14}$ & 3.7$\times$10$^{-58}$ \\
k$_{15}$ & 1.2$\times$10$^{-57}$ \\
k$_{16}$ & 8.4$\times$10$^{-11}$ \\
k$_{-16}$ & 9.5$\times$10$^{-68}$ \\
\hline
\end{tabular}
\end{table}

The theoretical value of k$_h$ is equal to the value of k$_{16}$. Thus the first step is the rate-limiting step. The theoretical value of k$_g$ is nearly the same as the value of k$_{14}$, however because the reverse rate coefficient (k$_{-14}$) is comparable to the rate coefficient of the second step (k$_{15}$), the value of k$_g$ is slightly smaller than that of k$_{11}$. We calculate k$_g$ to be 4.3$\times$10$^{-10}$ cm$^{3}$s$^{-1}$, which is within the range of experimental values for \ce{^3CH2 + H -> CH + H2}. We calculate k$_h$ to be 8.4$\times$10$^{-11}$ cm$^{3}$s$^{-1}$, which is less than a factor of 2 larger than the suggested value by Tsang and Hampson\cite{Reference509}. Our calculations show \ce{CH3 -> CH + H2} is inefficient (k$_{15}$ = $\sim$10$^{-57}$ s$^{-1}$), therefore we do not consider the second step of this reaction in our study.

The same mechanistic approach can be used for the reverse reaction \ce{CH + H2 -> CH3* -> ^3CH2 + H}. This reaction could produce $^1$CH$_2$ + H as well, however k$_{-14}$ is $\sim$10 orders of magnitude larger than k$_{-16}$, suggesting the dominant pathway would be to produce $^3$CH$_2$ + H.

Several experiments have calculated the rate coefficient of \ce{CH + H2 -> products} by monitoring the decay of CH in the presence of H$_2$ \cite{Reference506,Reference516,Reference517,Reference518,Reference519,Reference520,Reference521,Reference575}. Becker et al.\cite{Reference517} find that 300 K is the threshold temperature, below which the CH$_3$ product is mainly formed, and above which the $^3$CH$_2$ + H products are mainly formed. Zabarnick et al.\cite{Reference506} also suggest CH$_3$ is the main product below temperatures of 300 K, and recommend a temperature of $>$ 400 K for the formation of CH$_2$ + H. 

Rate coefficients for the reaction of \ce{CH + H2 -> CH3} range from 1.0$\times$10$^{-12}$ to 1.6$\times$10$^{-10}$ cm$^3$s$^{-1}$.

Mayneris et al.\cite{Reference547} used the quasiclassical trajectory method to calculate the theoretical rate coefficient of \ce{CH + H2 -> CH3* -> $^3$CH2 + H}. They calculated a value of 3.5$\times$10$^{-11}$ cm$^3$s$^{-1}$.

We find the first step of the reverse reaction to be the rate-limiting step, i.e. \ce{CH + H2 -> CH3}. We calculate the rate coefficient of the reverse reaction to be 7.9$\times$10$^{-11}$ cm$^3$s$^{-1}$. This is within the range of experimental values.

We find the second step of the reverse reaction (\ce{CH3 -> CH2 + H}) to be too inefficient to consider in this network (i.e. k $<$10$^{-21}$ s$^{-1}$).

\subsubsection*{Case Study 6: \ce{CH2 + H2 -> CH3 + H}}

This reaction occurs on the singlet and triplet surfaces, as \ce{^1CH2 + H2 -> CH4* -> CH3 + H} and \ce{^3CH2 + H2 -> CH3 + H}, respectively.

Experiments have measured the rate of $^1$CH$_2$ decay or CH$_3$ production to calculate the rate coefficient of \ce{^1CH2 + H2 -> CH4* -> CH3 + H} at 295--298 K \cite{Reference510,Reference514,Reference515}. These values range from 7.0$\times$10$^{-12}$ to 1.3$\times$10$^{-10}$ cm$^{3}$s$^{-1}$. Studies reviewing these experiments suggest a rate coefficient of 1.2$\times$10$^{-10}$ cm$^{3}$s$^{-1}$\cite{Reference509,Reference451}.

We find no theoretical rate coefficients for the \ce{^1CH2 + H2 -> CH4* -> CH3 + H} reaction.

We find the first step to be the rate-limiting step in the reaction \ce{^1CH2 + H2 -> CH4* -> CH3 + H}. We calculate the rate coefficient to be 1.0$\times$10$^{-11}$ cm$^{3}$s$^{-1}$, which is within the range of experimental values. Because CH$_4$ is a stable product in our reaction network, we only include the first step of this reaction in our network. The second step, \ce{CH4 -> CH3 + H}, is very inefficient (k $\sim$ 10$^{-60}$ s$^{-1}$) and we do not include it in our network.

Although no experiments have directly measured the rate coefficient of \ce{^3CH2 + H2 -> CH3 + H}, a few models have place an upper bound on its value by considering the affect of various gases on the $^3$CH$_2$ molecule. These upper bounds range from 5.0$\times$10$^{-15}$ to 5.0$\times$10$^{-14}$ cm$^{3}$s$^{-1}$.

Lu et al.\cite{Reference546} calculated the theoretical rate coefficient of \ce{^3CH2 + H2 -> CH3 + H} using transition state theory. They employed the G2M(RCC2) computational method with B3LYP optimized geometries and obtained a value of 1.5$\times$10$^{-18} $cm$^{3}$s$^{-1}$. 

We calculate the rate coefficient of \ce{^3CH2 + H2 -> CH3 + H} to be 2.5$\times$10$^{-16}$ cm$^{3}$s$^{-1}$. This value agrees with the upper bounds for the rate coefficient from experiments.

We find the reverse rate coefficient, \ce{CH3 + H -> ^3CH2 + H2} to have a value of 1.4$\times$10$^{-20}$ cm$^{3}$s$^{-1}$. We include this reverse reaction in our network as its rate coefficient is within the threshold of what we define to be a fast reaction (i.e. k $>$10$^{-21}$ cm$^{3}$s$^{-1}$).

\subsubsection*{Case Study 7: \ce{CH2 + N -> HCN + H}}

Catling and Kasting\cite{2007AsNow..22e..76R} suggest \ce{CH2 + N -> HCN + H} is one of the main pathways forming HCN in the early atmosphere. They note however that the rate constant for this reaction has not yet been studied experimentally.

Herbst et al.\cite{Reference574} performed quantum dynamics simulations to calculate the rate coefficient of \ce{^3CH2 + ^4N -> H2CN* -> HCN + H}. They calculated a value of 7.9$\times$10$^{-11}$ cm$^{3}$s$^{-1}$.

We find no direct reaction pathway for \ce{CH2 + N -> HCN + H} on the doublet, quartet, or sextet PES's. We do however find two-step reactions \ce{CH2 + N -> H2CN} and \ce{H2CN -> HCN + H} on the doublet and quartet surfaces.

We list the calculated reaction rate coefficients on the doublet and quartet energy surfaces in Table~\ref{Table6}.

\begin{table}[t]
\centering
\caption{Calculated rate coefficients for \ce{CH2 + N -> H2CN}, and \ce{H2CN -> HCN + H} on the doublet and quartet potential energy surfaces. In all simulations, the BHandHLYP method was used with the aug-cc-pVDZ basis set. First-order rate coefficients have units s$^{-1}$. Second-order rate coefficients have units cm$^{3}$s$^{-1}$. \label{Table6}}
\begin{tabular}{ccc}
\\
\multicolumn{1}{c}{Reaction} &
\multicolumn{1}{c}{\hspace{0.3cm}k(298) doublet} &
\multicolumn{1}{c}{\hspace{0.3cm}k(298) quartet}\\ \hline \\[-2mm]
\ce{^1CH2 + ^2N -> H2CN} & 1.5$\times$10$^{-10}$ & \\[+1mm]
\ce{^1CH2 + ^4N -> H2CN} & & 1.1$\times$10$^{-10}$ \\[+1mm]
\ce{^3CH2 + ^4N -> H2CN} & 1.3$\times$10$^{-10}$ & \\[+1mm]
\ce{^3CH2 + ^2N -> H2CN} & 2.7$\times$10$^{-10}$ & 4.3$\times$10$^{-10}$ \\[+1mm]
\ce{H2CN -> HCN + H} & 1.6$\times$10$^{-11}$ & 4.6$\times$10$^{-24}$ \\
\hline
\end{tabular}
\end{table}

All spin configurations of \ce{CH2 + N -> H2CN} are barrierless and have efficient reaction rate coefficients. \ce{H2CN -> HCN + H}, however, is only efficient on the doublet surface. We distinguish between the quartet and doublet H$_2$CN molecules in our network, as the deexcitation of $^4$H$_2$CN to $^2$H$_2$CN is spin-forbidden, and we can't assume $^4$H$_2$CN will efficiently decay into its ground state in an atmosphere.

Our calculated rate coefficient for \ce{^3CH2 + ^4N -> H2CN} is approximately a factor of 1.5 greater than the previous theoretical value \cite{Reference574}.

We include all five \ce{CH2 + N -> H2CN} reaction spin configurations as well as the efficient doublet \ce{H2CN -> HCN + H} reaction in our network.

\subsubsection*{Case Study 8: \ce{2CH2 -> C2H4* -> products}}

There are three spin configurations for this reaction on a total of two PES's. On the singlet surface, there is \ce{^3CH2 + ^3CH2 -> ^1C2H4* -> C2H3* + H* -> C2H2 + 2H} and \ce{^1CH2 + ^1CH2 -> ^1C2H4* -> products}, and on the triplet PES there is \ce{^3CH2 + ^1CH2 -> ^3C2H4* -> products}.

Braun et al.\cite{Reference510} monitored the decay of $^3$CH$_2$ and the production of C$_2$H$_2$ in experiments to measure the rate coefficient of \ce{^3CH2 + ^3CH2 -> C2H2 + product} at 298 K. This measurement was 5.3$\times$10$^{-11}$ cm$^3$s$^{-1}$. Braun et al.\cite{Reference510} assumed that molecular hydrogen was produced along with C$_2$H$_2$ in this reaction, however Becerra et al.\cite{Reference541} modeled the reaction network starting from the decomposition of ketene and found that \ce{^3CH2 + ^3CH2 -> C2H2 + 2H} was more likely. Becerra et al.\cite{Reference541} found that \ce{^3CH2 + H -> CH + H2} can account for the molecular hydrogen observed in reactions of this kind.

Braun et al.\cite{Reference510} suggest that the reaction of $^3$CH$_2$ with $^3$CH$_2$ passes through the C$_2$H$_4$ intermediate. 

Jasper et al.\cite{Reference577} calculate the theoretical rate coefficient for \ce{^3CH2 + ^3CH2 -> C2H4* -> C2H2 + 2H} using variable reaction coordinate transition state theory. Their value is 1.5 $\times$10$^{-10}$ cm$^3$s$^{-1}$.

There is no experimental data for \ce{^1CH2 + ^1CH2 -> C2H4* -> products}, however it is expected to proceed rapidly, and yield the same products as \ce{^3CH2 + ^3CH2 -> C2H4* -> C2H2 + 2H} \cite{Reference509}. Tsang and Hampson\cite{Reference509} recommend a value of 5.0$\times$10$^{-11}$ cm$^3$s$^{-1}$ for this reactions.

Similarly, there is no experimental data for \ce{^3CH2 + ^1CH2 -> ^3C2H4* -> products}, however it is also expected to be rapid. Conversely, it is suggested that the preferred products for this reaction are \ce{^3CH2 + ^1CH2 -> ^3C2H4* -> ^3C2H2 + H2}. Tsang and Hampson\cite{Reference509} suggest a value of 3.0$\times$10$^{-11}$ cm$^3$s$^{-1}$ for this reaction.

To our knowledge there have been no theoretical reaction rate coefficients for \ce{^1CH2 + ^1CH2 -> C2H2 + 2H}, or \ce{^1CH2 + ^3CH2 -> ^3C2H2 + H2} published to date.

Because in some of the other reactions in our network, C$_2$H$_4$ is a stable product, i.e. \ce{CH + CH4 -> C2H5* -> C2H4 + H}, \ce{CH2 + CH3 -> C2H5* -> C2H4 + H}, we only include the first steps of these reactions in our network (i.e.  \ce{2CH2 -> C2H4}). We find the first steps of reactions \ce{^3CH2 + ^3CH2 -> C2H4* -> C2H2 + 2H} and \ce{^1CH2 + ^1CH2 -> C2H4* -> C2H2 + 2H} to be the rate-limiting steps, and assume the same for \ce{^1CH2 + ^3CH2 -> ^3C2H4 -> products}.

We list the calculated reaction rate coefficients on the singlet and triplet energy surfaces in Table~\ref{Table8}.

\begin{table}[t]
\centering
\caption{Calculated rate coefficients for \ce{2CH2 -> C2H4} on the singlet and triplet potential energy surfaces. These rate coefficients are compared with the experimental rate coefficient of the multi-step reaction \ce{^3CH2 + ^3CH2 -> C2H4* -> C2H2 + 2H} as well as the suggested rate coefficients for \ce{^1CH2 + ^1CH2 -> C2H4* -> C2H2 + 2H}, and \ce{^1CH2 + ^3CH2 -> ^3C2H4* -> products} in the literature. We find the first steps of the reactions of \ce{^{3,1}CH2 + ^{3,1}CH2 -> C2H4* -> C2H2 + 2H} to be the rate-limiting steps and assume the same for \ce{^1CH2 + ^3CH2 -> ^3C2H4* -> products}. In all simulations, the BHandHLYP method was used with the aug-cc-pVDZ basis set. Rate coefficients have units cm$^{3}$s$^{-1}$. \label{Table8}}
\begin{tabular}{cccc}
\\
\multicolumn{1}{c}{} &
\multicolumn{1}{c}{$^3$CH$_2$ + $^3$CH$_2$} &
\multicolumn{1}{c}{\hspace{0.3cm}$^1$CH$_2$ + $^1$CH$_2$} &
\multicolumn{1}{c}{\hspace{0.3cm}$^1$CH$_2$ + $^3$CH$_2$}\\ \hline \\[-2mm]
k$_{calc}$(298) & 4.2$\times$10$^{-11}$ & 9.9$\times$10$^{-12}$ & $^a$3.5$\times$10$^{-11}$ \\[+1mm]
k$_{lit}$(298) & 5.3$\times$10$^{-11}$ & 5.0$\times$10$^{-11}$ & 3.0$\times$10$^{-11}$ \\[+1mm]
\hline
\multicolumn{4}{l}{\footnotesize $^a$ Simulations did not converge beyond a C-C bond distance of} \\
\multicolumn{4}{l}{\footnotesize 3.52$\AA$. Therefore the calculated rate coefficient is a lower bound.} \\
\end{tabular}
\end{table}

Our calculated k(298) value for \ce{^3CH2 + ^3CH2 -> C2H4} is within 30$\%$ of the experimental value for \ce{^3CH2 + ^3CH2 -> C2H2 + 2H}. The k(298) value for \ce{^1CH2 + ^1CH2 -> C2H4} is a factor of 5 smaller than the suggested value. Simulations did not converge for \ce{^1CH2 + ^3CH2 -> ^3C2H4} beyond a C-C reaction coordinate of 3.52 $\AA$, however using this location for the variational transition state leads to a calculated rate coefficient that is within 20$\%$ of the suggested value.

%therefore the Gibbs maximum was not found. However, the Gibbs maxima for \ce{^3CH2 + ^3CH2 -> C2H4} and \ce{^1CH2 + ^1CH2 -> C2H4} are at 3.3 and 3.79 $\AA$, respectively, therefore using the converged simulation at 3.52 $\AA$ gives us a fairly accurate lower bound, which is 

\subsubsection*{Case Study 9: \ce{CH2 + CH3 -> C2H5* -> C2H4 + H}}

This is a two step reaction, passing through the C$_2$H$_5$ intermediate \cite{Reference511,Reference512,Reference509}. On the doublet surface, both $^3$CH$_2$ and $^1$CH$_2$ can react with CH$_3$ to produce the C$_2$H$_5$ intermediate. On the quartet surface, $^3$CH$_2$ reacts with CH$_3$ to produce excited $^4$C$_2$H$_5$. However this reaction is higher in energy than the doublet reactions, and has a very slow rate coefficient ($\sim$10$^{-57}$ cm$^{3}$s$^{-1}$).

Pilling and Robertson\cite{Reference511} and Laufer and Bass\cite{Reference512} experimentally measured the production of various products (e.g. C$_2$H$_2$, C$_2$H$_4$, C$_2$H$_6$) to model the reaction network spanned by reactions between $^3$CH$_2$ and CH$_3$. Their models led to reaction rate coefficients of 5.0$\times$10$^{-11}$ and 1.0$\times$10$^{-10}$ cm$^{3}$s$^{-1}$, respectively. Wang and Fockenberg\cite{Reference513} performed similar experiments but used the $^3$CH$_2$ decay profile for their calculation, and obtained a rate of 2.1$\times$10$^{-10}$ cm$^{3}$s$^{-1}$ at 300 K. Deters et al.\cite{Reference578} measured the decay of both $^3$CH$_2$ and CH$_3$ in a similar experiment to obtain a rate coefficient of 1.1$\times$10$^{-10}$ cm$^{3}$s$^{-1}$ at 298 K. Baulch et al.\cite{Reference451} and Tsang and Hampson\cite{Reference509} reviewed various experiments and suggest a value of 7.0$\times$10$^{-11}$ cm$^{3}$s$^{-1}$ for the \ce{$^3$CH2 + CH3 -> C2H5* -> C2H4 + H} rate coefficient.

Conversely, there is no experimental data for \ce{^1CH2 + CH3 -> C2H5* -> C2H4 + H}, however the reaction is thought to proceed rapidly and suggested to have a rate coefficient near 3.0$\times$10$^{-11}$ cm$^{3}$s$^{-1}$ \cite{Reference509}.

There have been no published theoretical reaction rate coefficients for \ce{CH2 + CH3 -> C2H4 + H}.

\begin{figure}[!hbtp]
\centering
\includegraphics[width=\linewidth]{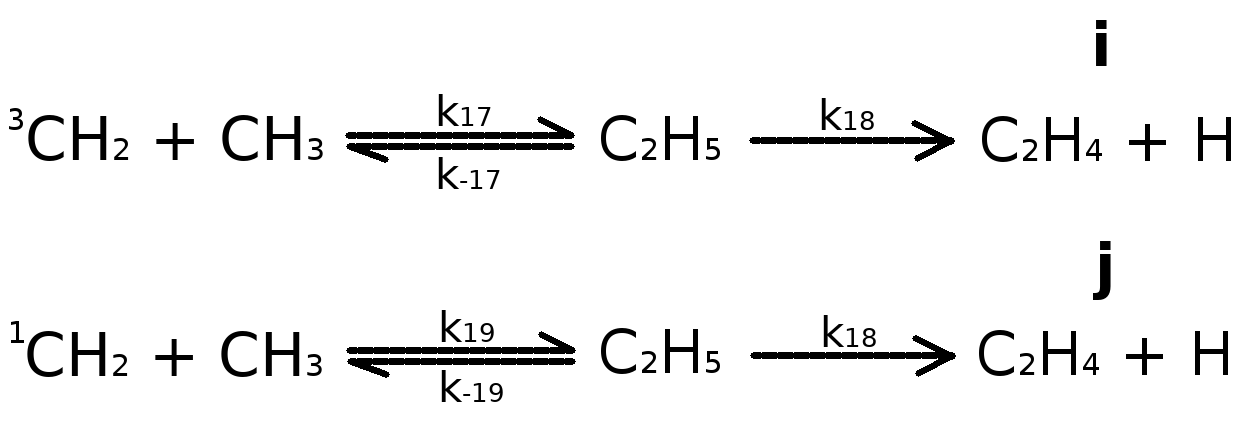}
\caption{Mechanistic model for the production of C$_2$H$_4$ + H from {\bf i)} $^3$CH$_2$ + CH$_3$ and {\bf j)} $^1$CH$_2$ + CH$_3$ on the doublet surface.}
\label{Figure6}
\end{figure}

We find the first steps of these reactions, \ce{^3CH2 + CH3 -> C2H5} and \ce{^1CH2 + CH3 -> C2H5}, do not have barriers. The second step however (\ce{C2H5 -> C2H4 + H}), has a barrier. The mechanistic model for the reaction, involving the triplet and singlet CH$_2$ molecule, is shown in Figure~\ref{Figure6}.

We calculate the upper bounds for the rate constants by assuming all C$_2$H$_5$ reacts to form C$_2$H$_4$ + H.

The steady-state solutions of the kinetic rate equations for this mechanistic model gives us the overall rate constant for \ce{CH2 + CH3 -> C2H4 + H}.

\begin{equation}
k_i = \frac{k_{18} k_{17}}{k_{-17} + k_{18}}
\end{equation}

\begin{equation}
k_j = \frac{k_{18} k_{19}}{k_{-19} + k_{18}}
\end{equation}

We list the calculated reaction rate coefficients for this mechanistic model in Table~\ref{Table7}.

\begin{table}[t]
\centering
\caption{Calculated overall rate coefficient for \ce{^3CH2 + CH3 -> C2H4 + H} and \ce{^1CH2 + CH3 -> C2H4 + H}, as well as the intermediate forward and reverse rate coefficients which were used in the calculation. In all simulations, the BHandHLYP method was used with the aug-cc-pVDZ basis set. Experiments at 298 K provide a k$_i$ value in the range of 5.0$\times$10$^{-11}$ to 2.1$\times$10$^{-10}$ cm$^{3}$s$^{-1}$ \cite{Reference511,Reference512,Reference513}. k$_j$ is suggested to have a rate coefficient near 3.0$\times$10$^{-11}$ cm$^{3}$s$^{-1}$ \cite{Reference509}. First-order rate coefficients have units s$^{-1}$. Second-order rate coefficients have units cm$^{3}$s$^{-1}$. \label{Table7}}
\begin{tabular}{ccc}
\\
\multicolumn{1}{c}{Rate coefficient} &
\multicolumn{1}{c}{k(298)}\\ \hline \\[-2mm]
k$_i$ & 8.8$\times$10$^{-12}$ \\[+1mm]
k$_j$ & 2.3$\times$10$^{-11}$ \\
k$_{17}$ & 8.8$\times$10$^{-12}$ \\
k$_{-17}$ & 1.4$\times$10$^{-48}$ \\
k$_{18}$ & 4.0$\times$10$^{-15}$ \\
k$_{19}$ & 2.3$\times$10$^{-11}$ \\
k$_{-19}$ & 3.0$\times$10$^{-57}$ \\
\hline
\end{tabular}
\end{table}

The theoretical values of k$_i$ and k$_j$ are equal to the values of k$_{17}$ and k$_{19}$, respectively. Thus the first steps of these reactions are the rate-limiting steps. At the BHandHLYP/aug-cc-pVDZ level of theory, we calculate k$_i$ to be 8.8$\times$10$^{-12}$ cm$^3$s$^{-1}$. This value is approximately a factor of 6 slower than the slowest experimental value. Similarly, we calculate k$_j$ to be 2.3$\times$10$^{-11}$ cm$^3$s$^{-1}$, which is within order unity of the suggested value.

\subsubsection*{Case Study 10: \ce{CH2 + CH4 -> 2CH3}}

This reaction occurs on the singlet PES as \ce{^1CH2 + CH4 -> C2H6* -> 2CH3}. and on the triplet PES as \ce{^3CH2 + CH4 -> 2CH3}.

Experimentalists have measured the rate coefficient of \ce{^1CH2 + CH4 -> C2H6* -> 2CH3} at 295--298 K by measuring the decay of $^1$CH$_2$ or the production of CH$_3$. These values range from 1.9$\times$10$^{-12}$ to 7.3$\times$10$^{-11}$ cm$^{3}$s$^{-1}$ \cite{Reference510,Reference514,Reference515}. Tsang and Hampson\cite{Reference509} reviewed these experiments and suggested a value of 7.1$\times$10$^{-11}$ cm$^{3}$s$^{-1}$. We find no published theoretical rate coefficients for this reaction.

We find the reaction on the singlet PES proceeds through the C$_2$H$_6$ intermediate and that the first step of this reaction is the rate-limiting step. We calculate its rate coefficient to be 6.1$\times$10$^{-13}$ cm$^{3}$s$^{-1}$, which is a factor of 3 smaller than the closest experimental value. Because C$_2$H$_6$ is a stable product in other reactions in our network (e.g. \ce{2CH3 -> C2H6}), we only consider the first step of this reaction in our network. The second step of this reaction \ce{C2H6 -> 2CH3} is too slow to consider in this network (k $\sim$ 10$^{-55}$ s$^{-1}$).

B{\"o}hland et al.\cite{Reference568} performed experiments on the reaction of $^3$CH$_2$ with n-hexane at T = 413--707 K to estimate the rate coefficient for  \ce{^3CH2 + CH4 -> 2CH3} at 298 K. They calculated a value of 3.1$\times$10$^{-19}$ cm$^{3}$s$^{-1}$. Braun et al.\cite{Reference510} placed an upper bound on the rate coefficient by considering the affect of various gases on the $^3$CH$_2$ molecule. They estimate a value of $<$3.1$\times$10$^{-14}$ cm$^{3}$s$^{-1}$. Tsang and Hampson\cite{Reference509} suggest an upper bound of 3.0$\times$10$^{-19}$ cm$^{3}$s$^{-1}$ based on the results of a photolysis study of the CH$_2$CO-CH$_4$ system. We find no theoretical rate coefficients for the \ce{^3CH2 + CH4 -> 2CH3} reaction.

We calculate \ce{^3CH2 + CH4 -> 2CH3} to have a rate coefficient of 1.4$\times$10$^{-16}$ cm$^{3}$s$^{-1}$. This value agrees with the upper bound from Braun et al.\cite{Reference510}, and is a couple orders of magnitude higher than the experimental value from B{\"o}hland et al.\cite{Reference568}.

\subsubsection*{Case Study 11: \ce{CH + N -> HCN -> CN + H}}

There are two spin configurations for this reaction on the triplet surface, which pass through the excited $^3$HCN intermediate: \ce{CH + ^4N -> ^3HCN -> CN + H} and \ce{CH + ^2N -> ^3HCN -> CN + H}. There is also potentially a reaction of CH + $^2$N on the singlet surface to produce ground state HCN, however we were unable to obtain a convergent solution for such a reaction. Moreover, there is no experimental or past theoretical work for a singlet surface reaction of CH + $^2$N to suggest it occurs efficiently. For these reasons, we only consider the two spin configurations on the triplet surface in this network.

A few experiments have measured the rate coefficient of \ce{CH + ^4N -> CN + H} at 296--298 K by monitoring the decay of CH and/or the production of CN \cite{Reference570,Reference571,Reference572}. The experimental values of the rate coefficient range from 2.1$\times$10$^{-11}$ to 1.6$\times$10$^{-10}$ cm$^{3}$s$^{-1}$.

Daranlot et al.\cite{Reference572} performed quantum dynamics calculations to obtain a theoretical rate coefficient for \ce{CH + ^4N -> ^3HCN -> CN + H}. They calculate a value of k(298) = 1.2$\times$10$^{-10}$ cm$^{3}$s$^{-1}$.

We find no experimental or theoretical rate coefficients for \ce{CH + ^2N -> ^3HCN -> CN + H}.

Our theoretical calculations show both \ce{CH + ^4N} and \ce{CH + ^2N} react without a barrier to form the $^3$HCN intermediate. We find the first step for both of these reactions to be the rate-limiting steps. The second step of this reaction, i.e., the decay of $^3$HCN into CN + H, is extremely efficient (k = 3.6$\times$10$^{9}$ s$^{-1}$).

We calculate the rate coefficient for \ce{CH + ^4N -> ^3HCN -> CN + H} to be 1.1$\times$10$^{-10}$ cm$^{3}$s$^{-1}$. This value is within the range of experimental values, and agrees well with the previous calculated theoretical value \cite{Reference572}.

We calculate the rate coefficient for \ce{CH + ^2N -> ^3HCN -> CN + H} to be 2.7$\times$10$^{-10}$ cm$^{3}$s$^{-1}$.

\subsubsection*{Case Study 12: \ce{CH + CH4 -> C2H5* -> C2H4 + H}}

This reaction occurs on the doublet surface. Several experiments have calculated the rate coefficient for this reaction at 295--298 K by monitoring the production of C$_2$H$_4$ or the decay of CH \cite{Reference519,Reference520,Reference521,Reference523,1997AA...323..644C,Reference561,Reference562,Reference563,Reference566}. The experimental rate coefficient ranges from 2.0$\times$10$^{-12}$ to 3.0$\times$10$^{-10}$ cm$^{3}$s$^{-1}$.

A pair of theoretical studies have been performed on this reaction, however theoretical rate coefficients were not calculated \cite{Reference564,Reference565}.

At the BHandHLYP/aug-cc-pVDZ level of theory, we find this reaction to have a small barrier (E$_0$ = 11.5 kJ mol$^{-1}$). This is smaller than the barrier predicted by Yu et al.\cite{Reference564} (E$_0$ = 57.3 kJ mol$^{-1}$), who used the M{\/o}ller-Plesset perturbation theory (MP) method with geometries optimized using the Hartree-Fock method. However, Wang et al.\cite{Reference565} calculated the reaction to be barrierless (-1.3 kJ mol$^{-1}$) using the MP method with MP optimized geometries. Experiments suggests the reaction is barrierless, with an activation energy of -1.7 kJ mol$^{-1}$ \cite{Reference561}. At the B3LYP/aug-cc-pVDZ level of theory, we find this reaction to be barrierless, with an activation energy of -18.2 kJ mol$^{-1}$. Because experiment predicts this reaction to be barrierless \cite{Reference561,Reference563,Reference566}, and the existence of the theoretical barrier is dependent on the computational method, we artificially remove the barrier from our calculation of the rate coefficient at the BHandHLYP/aug-cc-pVDZ level of theory.

We find the first step of this reaction \ce{CH + CH4 -> C2H5} to be the rate-limiting step, with a barrierless rate coefficient of 3.8$\times$10$^{-13}$ cm$^{3}$s$^{-1}$. This is a factor of 5 slower than the nearest experimental value. We calculate the rate coefficient of second step of this reaction \ce{C2H5 -> C2H4 + H} to be 1.8$\times$10$^{-11}$ s$^{-1}$, suggesting the C$_2$H$_5$ intermediate is fairly unstable. Thus we include this two-step reaction in our network as a single step \ce{CH + CH4 -> C2H4 + H}.

\subsubsection*{Case Study 13: \ce{NH + H <-> N + H2}}

This reaction has two spin configurations on the doublet PES, \ce{^1NH + H -> ^2N + H2} and \ce{^3NH + H -> ^2N + H2}, and one spin configuration on the quartet PES, \ce{^3NH + H -> ^4N + H2}.

Adam et al.\cite{Reference527} calculated the experimental rate coefficient of \ce{^3NH + H -> ^4N + H2} at 298 K by monitoring the decay of $^3$NH. They found the rate coefficient to have a value of 3.2$\times$10$^{-12}$ cm$^{3}$s$^{-1}$.

Adam et al.\cite{Reference527} also used the classical trajectory method to calculate the theoretical rate coefficient for \ce{^3NH + H -> ^4N + H2} at the MRCI/aug-cc-pVQZ level of theory. They found this reaction proceeds directly, rather than through the NH$_3$ intermediate. They calculated the rate coefficient to be 1.5$\times$10$^{-12}$ cm$^{3}$s$^{-1}$. Other theoretical works calculated the rate coefficient with CVT and QCT to range from 2.0$\times$10$^{-13}$ to 5.2$\times$10$^{-12}$ cm$^{3}$s$^{-1}$ \cite{Reference528,Reference585}.

We find no published experimental or theoretical rate coefficients for the two spin configurations on the doublet PES.

On the quartet surface, we calculate the \ce{^3NH + H -> H2 + ^4N} configuration to be 1.4$\times$10$^{-11}$ cm$^{3}$s$^{-1}$. This is a factor of 4 greater than the experimental value reported by Adam et al.\cite{Reference527}.

On the doublet surface, we do not calculate the \ce{^1NH + H -> H2 + ^2N} configuration as $^1$NH is not efficiently produced in this reaction network. 

We find the \ce{^3NH + H -> H2 + ^2N} configuration to proceed through the NH$_2$ intermediate. This is consistent with theoretical studies of the reverse reaction \cite{Reference551,Reference567}. We find the total forward rate coefficient to be too slow to consider in this study ($\sim$10$^{-80}$ cm$^{3}$s$^{-1}$).

Regarding the reverse reaction, various experiments have been performed on the deactivation of $^2$N by H$_2$ at 295--300 K \cite{Reference548,Reference552,Reference553,Reference554,Reference555,Reference556,Reference557,Reference558,Reference559}. The rate coefficients have been measured by monitoring the decay of $^2$N and range from 1.7--5.0$\times$10$^{-12}$ cm$^{3}$s$^{-1}$. Donovan and Husain\cite{Reference560} indicate that $^2$N + H$_2$ should readily undergo chemical reaction into $^3$NH + H via a direct path on the doublet PES. However, theoretical works suggest this reaction will proceed through the NH$_2$ intermediate \cite{Reference551,Reference567}.
Herron\cite{Reference549} reviewed the deactivation experiments and suggested a rate coefficient of 2.2$\times$10$^{-12}$ cm$^{3}$s$^{-1}$ for \ce{^2N + H2 -> ^3NH + H}.

Theoretical rate coefficient calculations of the reaction \ce{^2N + H2 -> ^3NH + H} have been performed using QCT \cite{Reference551,Reference550,Reference567}, quantum dynamics \cite{Reference550}, and CVT \cite{Reference551} with the CASSCF and MRCI computational methods. Kobayashi et al.\cite{Reference551} and Pederson et al.\cite{Reference567} suggest this reaction proceeds through the NH$_2$ intermediate. Pederson et al.\cite{Reference567} find the H$_2$ molecule approaches the N atom perpendicularly, and that there is no collinear reaction path. The calculated theoretical rate coefficients range from 8.9$\times$10$^{-13}$--3.3$\times$10$^{-12}$ cm$^{3}$s$^{-1}$.

Experimental and theoretical studies both suggest \ce{^2N + H2 -> ^3NH + H} has a small energy barrier. The experimental value is 7.3 kJ mol$^-1$ \cite{Reference552}.

We find no published experimental or theoretical rate coefficients for the two other reverse reaction spin configurations (\ce{^2N + H2 -> ^1NH + H} and \ce{^4N + H2 -> ^3NH + H}). 

At the BHandHLYP/aug-cc-pVDZ level of theory, we find the first step of the reverse reaction, \ce{H2 + ^2N -> NH2}, to be barrierless. This step is also the rate-limiting step. Similarly to Pederson et al.\cite{Reference567}, we find the H$_2$ molecule approaches the N atom perpendicularly. The second step, \ce{NH2 -> 3NH + H}, proceeds through a barrier. We calculate the overall barrierless rate coefficient to be 9.7$\times$10$^{-10}$ cm$^{3}$s$^{-1}$. This value is over 2 orders of magnitude larger than the experimental values. This disagreement with experiment is due to the lack of a barrier calculated when using the BHandHLYP method. For this reason, we introduce the experimental barrier of 7.3 kJ mol$^{-1}$ \cite{Reference552} to our calculation to obtain an overall rate coefficient of 5.1 $\times$10$^{-10}$ cm$^{3}$s$^{-1}$. This value is only 1 order of magnitude larger than the experimental value. 

We expect the remaining discrepancy to be a result of our chosen computational method, as our reaction geometry is the same as other theoretical works \cite{Reference551,Reference567}.

We find the other two reverse rate coefficients (\ce{^2N + H2 -> ^1NH + H} and \ce{^4N + H2 -> ^3NH + H}) to be too inefficient to consider in this study (k $<$ 10$^{-21}$ cm$^{3}$s$^{-1}$).

\subsubsection*{Case Study 14: \ce{NH + N -> N2H* -> N2 + H}}

This reaction occurs on the doublet PES. There are three possible spin configurations: \ce{^3NH + ^4N -> N2H* -> N2 + H}, \ce{^3NH + ^2N -> N2H* -> N2 + H}, and \ce{^1NH + ^2N -> N2H* -> N2 + H}. Because $^1$NH is not produced efficiently by any reaction in this study, we only analyze the two spin configurations involving $^3$NH.

Hack et al.\cite{Reference525} experimentally measured the rate coefficient of \ce{^3NH + N -> products} at 298 K by monitoring the decay profile of $^3$NH. They measured the value to be 2.5$\times$10$^{-11}$ cm$^{3}$s$^{-1}$.

Konnov and De Ruyck\cite{Reference526} used the experimental value from Hack et al.\cite{Reference525}, as well as a suggested T$^{0.5}$ dependence to estimate a value of 2.6$\times$10$^{-11}$ cm$^{3}$s$^{-1}$.

Caridade et al.\cite{Reference524} calculated the theoretical rate coefficient of \ce{^3NH + ^4N -> N2H* -> N2 + H} to be 1.9$\times$10$^{-11}$ cm$^{3}$s$^{-1}$ using quasi-classical trajectory theory at the MRCI/aug-cc-pVQZ level of theory.

We find no published experimental or theoretical rate coefficients for \ce{^3NH + ^2N -> N2H* -> N2 + H}.

Consistent with a previous theoretical study, we find the \ce{^3NH + ^2N} reaction proceeds through the N$_2$H intermediate. We find the rate-limiting step to be \ce{^3NH + ^4N -> N2H}, with a rate coefficient of 
4.0$\times$10$^{-11}$ cm$^{3}$s$^{-1}$. This value is only a factor of 1.5 larger than the experimental value, and a factor of 2 larger than the theoretical value.

\begin{table}[t]
\centering
\caption{Reaction path symmetry numbers for each reaction ($\sigma$), as well as the rotational symmetry numbers of the reactants ($\sigma_i$) and transition states ($\sigma^{\dagger}$) used in the calculation. All steps in multi-step reactions are included. Spins are labeled only if reaction spin configurations have different reaction path symmetry numbers. $\sigma$ = $\frac{\prod_{i=1}^{N} \sigma_i}{\sigma^{\dagger}}$. \label{TableE}}
\begin{tabular}{ccccc}
\\
\multicolumn{1}{c}{Reaction Equation} &
\multicolumn{1}{c}{$\sigma_1$} &
\multicolumn{1}{c}{$\sigma_2$} &
\multicolumn{1}{c}{$\sigma^{\dagger}$} &
\multicolumn{1}{c}{$\sigma$}\\ \hline \\[-2mm]
\ce{H2CN -> HCN + H} & 2 &  & 1 & 2\\[+1mm]
\ce{HCN + H -> H2CN} & 1 & 1 & 1 & 1\\
\ce{H2CN + H -> HCN + H2} & 2 & 1 & 1 & 2\\
\ce{H2CN + N -> HCN + NH} & 2 & 1 & 1 & 2\\
\ce{2H2CN -> HCN + H2CNH} & 2 & 2 & 1 & 4\\
\ce{CH4 + H -> CH3 + H2} & 12 & 1 & 3 & 4 \\
\ce{CH4 + N -> H3CNH} & 12 & 1 & 1 & 12 \\
\ce{H3CNH -> H2CNH + H} & 1 & 1 & 1 & 1 \\
\ce{CH3 + H -> CH4} & 6 & 1 & 3 & 2 \\
\ce{CH3 + H2 -> CH4 + H} & 6 & 2 & 3 & 4 \\
\ce{CH3 + N -> H3CN} & 6 & 1 & 3 & 2\\
\ce{H3CN -> H2CN + H} & 3 & & 1 & 3\\
\ce{H3CN -> H2CNH} & 3 & & 1 & 3\\
\ce{H2CNH -> H2CN + H} & 1 & & 1 & 1\\
\ce{2CH3 -> C2H6} & 6 & 6 & 6 & 6 \\
\ce{2CH3 -> CH2 + CH4} & 6 & 6 & 1 & 36 \\
\ce{CH2 + H -> CH3} & 2 & 1 & 1 & 2\\
\ce{CH2 + H2 -> CH4} & 2 & 2 & 1 & 4\\
\ce{CH2 + H2 -> CH3 + H} & 2 & 2 & 2 & 2\\
\ce{CH2 + N -> H2CN} & 2 & 1 & 1 & 2\\
\ce{^1CH2 + ^1CH2 -> C2H4} & 2 & 2 & 1 & 4\\
\ce{^1CH2 + ^3CH2 -> C2H4} & 2 & 2 & 1 & 4\\
\ce{^3CH2 + ^3CH2 -> C2H4} & 2 & 2 & 2 & 2\\
\ce{CH2 + CH3 -> C2H5} & 2 & 6 & 1 & 12 \\ 
\ce{C2H5 -> C2H4 + H} & 1 & & 1 & 1\\
\ce{CH2 + CH4 -> C2H6} & 2 & 12 & 1 & 24\\
\ce{CH2 + CH4 -> 2CH3} & 2 & 12 & 1 & 24\\
\ce{CH + H2 -> CH3} & 1 & 2 & 1 & 2\\
\ce{CH + N -> HCN} & 1 & 1 & 1 & 1\\
\ce{HCN -> CN + H} & 1 &  & 1 & 1\\
\ce{2CH -> C2H2} & 1 & 1 & 1 & 1 \\
\ce{CH + CH4 -> C2H5} & 1 & 12 & 1 & 12 \\
\ce{NH + H -> H2 + N} & 1 & 1 & 1 & 1 \\
\ce{N + H2 -> NH2} & 1 & 2 & 2 & 1 \\
\ce{NH2 -> NH + H} & 2 &  & 1 & 2 \\
\ce{NH + N -> N2H} & 1 & 1 & 1 & 1 \\
\ce{N2H -> N2 + H} & 1 &  & 1 & 1 \\
\hline
\end{tabular}
\end{table}

\subsection*{Reaction Path Symmetry Numbers}

The reaction path symmetry number, or reaction path multiplicity, can be calculated with the following equation.

\begin{equation}
\sigma = \frac{\prod_{i=1}^{N} \sigma_i}{\sigma^{\dagger}}
\end{equation}
where $\sigma$ is the reaction path symmetry number, $\sigma_i$ is the rotational symmetry number of reactant $i$, and $\sigma^{\dagger}$ is the rotational symmetry number of the transition state.

In Table~\ref{TableE} we list the reaction path symmetry numbers for all the reactions in our network, as well as the rotational symmetry numbers of the reactants and products used in the calculation.

\end{document}